\documentclass[reprint,amsmath,amssymb,aps]{revtex4-2}

\usepackage{graphicx}
\usepackage{dcolumn}
\usepackage{bm}
\usepackage{float}
\usepackage{xcolor}
\usepackage{mathrsfs}
\usepackage{tikz} 

\begin{document}

\title{Gap Anisotropy in Layered Superconductors Due to Rashba and Dresselhaus Spin-Orbit Interactions}

\author{Bahruz Suleymanli}
\affiliation{Physics Department, Yıldız Technical University, 34220 Esenler, Istanbul, Türkiye}

\author{B. Tanatar}
\affiliation{Department of Physics, Bilkent University, 06800 Ankara, Türkiye}

\date{\today}

\begin{abstract}
The theory of layered superconductors is extended in the presence of Rashba and Dresselhaus spin-orbit interactions (SOIs). Using the intralayer BCS-like pairing interaction and employing the Gor’kov formalism, we obtain analytical expressions for the temperature Green’s functions and determine the gap function $\Delta$ which becomes complex in the presence of SOIs.
In the absence of SOIs, $\Delta$ is isotropic at both zero and finite temperatures, but it becomes anisotropic even in the presence of a single SOI. This anisotropy is related to the extra $\cos k_z$ factors in which the $k_z$ momentum along the $z$ direction contributes to the magnitude of the gap function. It is also found that SOIs suppress $\Delta$ at both zero and finite temperatures, and for certain critical values of SOIs and beyond $\Delta$ vanishes. Analytical expressions for the critical values of SOIs at zero temperature are obtained. Additionally, how the BCS equation for layered superconductors changes in the presence of SOIs is determined. 
\end{abstract}

\maketitle

\section{\label{sec:1}Introduction}

Rashba and Dresselhaus SOIs are foundational in the development of spintronic technologies due to their capacity to control spin dynamics via electric fields in systems lacking inversion symmetry \cite{PhysRevLett.90.076807, Meier2007, Manchon2015, Bercioux_2015, Bihlmayer2022}. The Rashba effect, with spin splitting linearly dependent on momentum, is prominent in structures with structural asymmetry, such as the Au(111) surface. This effect is observable through angle-resolved photoemission spectroscopy (ARPES), which highlights the alignment of electron spin with a momentum-dependent effective magnetic field, a feature advantageous for external manipulation via electric and magnetic fields \cite{PhysRevB.69.241401, Mende2022}. Conversely, Dresselhaus SOI arises from bulk inversion asymmetry in crystals, notably zinc-blende structures like GaAs, introducing cubic spin splitting terms \cite{PhysRev.100.580}. These interactions are particularly useful in heterostructures, where controlled tuning can reduce spin relaxation, thereby supporting long-lived spin states for advanced spintronic devices \cite{PhysRevB.84.075326, PhysRevMaterials.3.084416, PhysRevB.107.035155}. Experimental studies on 2D transition metal dichalcogenides and complex heterostructures highlight precise control over spin splitting and spin textures, which has fueled advancements in high-performance spintronics \cite{Wang2012, PhysRevLett.112.156404}.

In superconductors, Rashba and Dresselhaus SOIs significantly alter spin dynamics and electronic structure, with profound implications for topological superconductivity \cite{Sato_2017, Sharma_2022}. These interactions induce spin-split bands that can enhance superconductivity in noncentrosymmetric materials, creating conditions favorable for Majorana fermions and robust spin textures, which are critical for topological quantum computing \cite{PhysRevLett.127.237601, Bihlmayer2022}. Specifically, the Rashba SOI facilitates topological phase transitions in the presence of a Zeeman field, augmenting nondegenerate spin structures such that conventional s-wave superconductivity can exhibit p-wave characteristics \cite{PhysRevLett.103.020401, PhysRevLett.104.040502, PhysRevB.97.195421}. Experimental results on proximity-induced superconductivity in topological insulators like HgTe and InSb nanowires have shown zero-bias conductance peaks, which are indicative of Majorana zero modes \cite{Mourik2012, Deng2012, PhysRevResearch.2.013377}. Significant experimental milestones include the observation of 4$\pi$-periodic Josephson supercurrent in HgTe and the demonstration of a robust superconducting gap in InAs nanowires with epitaxial Al coating, both environments conducive to Majorana phenomena \cite{PhysRevX.7.021011, Albrecht2016}. Additionally, atomic chains on Pb surfaces with strong SOI offer further experimental evidence of Majorana zero modes localized at chain edges, marking an essential step toward practical topological quantum computing \cite{Stevan2014}.

Layered systems provide a distinct platform for studying Rashba-type SOI effects in atomic-layer superconductors \cite{Yoshizawa2021, Uchihashi2021, PhysRevB.103.245113, SAKAMOTO2022100665}. A prominent example involves Rashba SOI on a Si(111)-($\sqrt{7} \times \sqrt{3}$)-In surface, where experiments revealed an in-plane upper critical magnetic field exceeding the Pauli limit by nearly three times at zero temperature \cite{Yoshizawa2021}. This phenomenon, enabled by dynamic spin-momentum locking, was confirmed through in-situ electron transport measurements and density functional theory (DFT), highlighting the resilience of superconductivity under high magnetic fields. Such spin-momentum locking forces electron spins to flip at each scattering event, suppressing paramagnetic pair-breaking and preserving superconductivity. Furthermore, Rashba SOI has been shown to induce spin-splitting in materials like Ge(111)-$\beta$($\sqrt{3} \times \sqrt{3}$)-Pb and Si(111)-($\sqrt{3} \times \sqrt{3}$)-TlPb, as observed through ARPES \cite{yaji2010large, Uchihashi2021}. In these systems, Rashba SOI-stabilized spin textures support unusual phenomena, such as spatially modulated superconducting order parameters, suggesting atomic-layer superconductors as ideal platforms for realizing unconventional superconductivity and exploring topological states.

Studies on a one-atom-layer Tl-Pb compound on Si(111) have demonstrated a large Rashba spin splitting ($\approx$ 250 meV) and two-dimensional superconductivity with a transition temperature of 2.25 K \cite{PhysRevLett.115.147003}. Observations from ARPES and transport measurements in this system reveal strong electron-phonon coupling and a Berezinskii-Kosterlitz-Thouless (BKT) transition, both of which reflect its two-dimensional superconducting nature. Here, Rashba SOI promotes mixed spin-singlet and spin-triplet pairing, with a potential for chiral p-wave topological superconducting states. Similar unconventional superconducting phases, such as complex-stripe (CS) and helical phases, have been observed in multilayer Pb films, where Rashba SOI induces spatially modulated order parameters, extending the exploration of topological superconductivity in systems lacking inversion symmetry \cite{PhysRevB.101.184502}.

The experimental studies mentioned above, along with ongoing experiments, emphasize the importance of analyzing Rashba and Dresselhaus SOI effects in layered superconductors. In this context, it is theoretically reasonable to model these superconductors as periodically arranged layers, where Cooper pairs can tunnel between the superconducting planes \cite{bulaevskii1973inhomogeneous, Klemm1975, Gulacsi1988, Schneider1989, EPNakhmedov_1994}. This approach also extends the analysis of these SOI effects to high-temperature superconducting crystals, where the strong anisotropy in both structural and physical properties makes the combined study of SOI and anisotropy effects particularly intriguing \cite{Wesche2025}. Notably, in almost all these crystals, an energy gap anisotropy has been observed experimentally, supporting the unique properties of high-temperature superconductors \cite{PhysRevLett.63.1008, PhysRevB.40.6902, PhysRevB.42.9973, PhysRevB.40.4478, PhysRevLett.70.1553, RevModPhys.77.721, RevModPhys.79.353, RevModPhys.95.031002}.

In this study, we extend the theory of layered superconductors to include the effects of Rashba and Dresselhaus SOIs. To this end, using the isotropic intralayer pairing interaction and employing the Gor’kov formalism, we derive the equations of motion incorporating temperature Green’s functions for layered superconductors, starting from the Hamiltonian of the problem examined in Sec.~\ref{sec:2}. In Sec.~\ref{sec:3}, we obtain analytical expressions for the temperature Green’s functions through Fourier transforms. These functions allow us to formulate the gap equation for the layered superconductors, as shown in Sec.~\ref{sec:4}. The results for zero and finite temperatures are provided in Secs.~\ref{sec:5} and \ref{sec:6}, respectively. Conclusions are given in Sec.~\ref{sec:levelc} where we also point out the relevance of our results to some recent experiments. 
Some technical details related to the calculations are presented in Appendices~\ref{app_1}, \ref{app_2}, \ref{app_3}, and \ref{app_4}.

\section{\label{sec:2}Model and Equations of motion}

The starting point of our theory is the Hamiltonian in the second quantization formalism for layered superconductors
\begin{equation}
    H = \sum_{j \sigma} \int d^2 r
        \left\{
                H_0 + H_T + H_\text{BCS} + H_\text{R} + H_\text{D}
        \right\},
\label{eq:main_H}    
\end{equation}
which includes free-electron propagation within each layer
\begin{equation}
    H_0 = - \psi_{j \sigma}^{\dagger}(\mathbf{r})
                 \frac{\mathbf{\nabla}^2}{2 m}
          \psi_{j \sigma}(\mathbf{r}), 
\label{eq:H_0}
\end{equation}
tunneling of the electrons between the layers
\begin{equation}
    H_T = \frac{J}{2}
          \left\{
                   \psi_{j \sigma}^{\dagger}(\mathbf{r}) \psi_{j+1, \sigma}(\mathbf{r}) + \text{H. c.}
          \right\},
\label{eq:H_T}    
\end{equation}
an intralayer BCS-type pairing interaction
\begin{equation}
    H_\text{BCS} = \frac{\lambda}{2} \psi_{j \sigma}^{\dagger}(\mathbf{r}) \psi_{j, - \sigma}^{\dagger}                     (\mathbf{r}) \psi_{j, -\sigma}(\mathbf{r}) \psi_{j \sigma}(\mathbf{r}),
\label{eq:H_BCS}    
\end{equation}
and the effects of the Rashba 
\begin{equation}
    H_\text{R} = \alpha \sum_{\gamma} \psi_{j \gamma}^{\dagger}(\mathbf{r})  
                 \left\{
                         \left(\sigma_x\right)_{\gamma \sigma} k_y - \left(\sigma_y\right)_{\gamma \sigma} k_x 
                 \right\}
                 \psi_{j \sigma}(\mathbf{r})
\label{eq:H_R}
\end{equation}
and Dresselhaus SOIs
\begin{equation}
    H_\text{D} = \beta \sum_{\gamma} \psi_{j \gamma}^{\dagger}(\mathbf{r})  
                 \left\{
                         \left(\sigma_x\right)_{\gamma \sigma} k_x - \left(\sigma_y\right)_{\gamma \sigma} k_y
                 \right\}
                 \psi_{j \sigma}(\mathbf{r}).
\label{eq:H_D}
\end{equation}
In the above, $\psi_{j \sigma}(\mathbf{r})$ and $\psi_{j \sigma}^\dagger(\mathbf{r})$ are the field operators in second quantization associated with an electron with spin $\sigma$ at position $\mathbf{r}$ in the $j$th layer, $J$ is the tunneling energy, $\lambda$ is the coupling constant of the BCS-type electron-electron interaction, $\alpha$ and $\beta$ are the strengths of the Rashba and Dresselhaus SOI, respectively, $\gamma$ is the spin index taking values $\uparrow$ and $\downarrow$, and $\sigma_{x,y}$ are the Pauli matrices in spin space. We note that the interlayer distance is uniformly set to 1. Additionally, in all our calculations, we use a system of units in which both Planck’s constant $\hbar$ and Boltzmann’s constant $k_B$ are equal to 1.

By writing the Heisenberg representation of the field operator $\psi_{j\sigma}(\mathbf{r})$ as $\psi_{j\sigma}(\mathbf{r}, \tau) = e^{\left(H-\mu N\right)\tau}\psi_{j\sigma}(\mathbf{r})e^{-\left(H-\mu N\right)\tau}$ in the equation of motion
\begin{equation}
    \frac{\partial}{\partial \tau} \psi_{j\sigma}(\mathbf{r}, \tau) = \left[H-\mu N, \psi_{j\sigma}                                                                                  (\mathbf{r}, \tau)\right],
\label{eq:eom_psi}    
\end{equation}
where $H$ is the Hamiltonian given by Eq.~(\ref{eq:main_H}) and $\mu$ is the chemical potential with $N = \int d^2r \psi_{j\sigma}^\dagger(\mathbf{r}) \psi_{j\sigma}(\mathbf{r})$ being the number of particles, and defining the temperature Green’s functions
\begin{subequations}
\begin{equation}
    G_{j j^{\prime}}^{a b}\left(\mathbf{r}, \tau; \mathbf{r}^{\prime}, \tau^\prime\right) 
    = -\left\langle T_\tau\left(\psi_{j a}(\mathbf{r}, \tau) \psi_{j^{\prime}                                   b}^{\dagger}\left(\mathbf{r}^{\prime}, \tau^\prime\right)\right)\right\rangle,
\label{eq:def_normal_green}    
\end{equation}
\begin{equation}
    F_{j j^{\prime}}^{a b}\left(\mathbf{r}, \tau; \mathbf{r}^{\prime}, \tau^\prime\right)
    = \left\langle T_\tau\left(\psi_{j a}(\mathbf{r}, \tau) \psi_{j^{\prime} b}\left(\mathbf{r}^{\prime},      \tau^\prime\right)\right)\right\rangle,
\label{eq:def_anomalous_green_1}     
\end{equation}
\begin{equation}
    F_{j j^{\prime}}^{\dagger a b}\left(\mathbf{r}, \tau; \mathbf{r}^{\prime}, \tau^\prime\right)
    = \left\langle T_\tau\left(\psi^\dagger_{j b}(\mathbf{r}, \tau) \psi^\dagger_{j^{\prime} a}\left(\mathbf{r}^{\prime},      \tau^\prime\right)\right)\right\rangle,
\label{eq:def_anomalous_green_2}     
\end{equation}
\end{subequations}
where Eqs.~(\ref{eq:def_anomalous_green_1}) and (\ref{eq:def_anomalous_green_2}) define the case with $a = -b$, and guided by the anticommutation relations of the field operators
we can find the following equation of motion containing the Green’s functions defined in Eqs.~(\ref{eq:def_normal_green}) -- (\ref{eq:def_anomalous_green_2})
\begin{widetext}
\begin{eqnarray}
        &&{\left[- \frac{\partial}{\partial \tau}+\frac{\mathbf{\nabla}^2}{2 m}+\mu\right] G^{ab}_{j j^{\prime}}\left(\mathbf{r}, \tau; \mathbf{r}^{\prime}, \tau^\prime\right)} -\frac{J}{2}\left\{G^{ab}_{j+1, j^{\prime}}\left(\mathbf{r}, \tau; \mathbf{r}^{\prime}, \tau^\prime\right)
        + G^{ab}_{j-1, j^{\prime}}\left(\mathbf{r}, \tau; \mathbf{r}^{\prime}, \tau^\prime\right)\right\} \nonumber\\
        &&+\lambda \sum_\gamma \left\{G_{jj}^{a\gamma}(0+) G_{jj^\prime}^{\gamma b}\left(\mathbf{r}, \tau; \mathbf{r}^{\prime}, \tau^\prime\right) 
        - G_{jj}^{\gamma\gamma}(0+) G_{jj^\prime}^{ab}\left(\mathbf{r}, \tau; \mathbf{r}^{\prime}, \tau^\prime\right) - F_{j j}^{a\gamma}\left(0+\right) F_{j j^{\prime}}^{\dagger\gamma b}\left(\mathbf{r}, \tau; \mathbf{r}^{\prime}, \tau^\prime\right)\right\}\nonumber\\
        && - \sum_\gamma \left\{\alpha \left(\left(\sigma_x\right)_{a \gamma} k_y - \left(\sigma_y\right)_{a \gamma} k_x \right) + \beta \left(\left(\sigma_x\right)_{a \gamma} k_x - \left(\sigma_y\right)_{a \gamma} k_y \right) \right\} G^{\gamma b}_{j j^{\prime}}\left(\mathbf{r}, \tau; \mathbf{r}^{\prime}, \tau^\prime\right) \nonumber\\
        &&=\delta_{j j^{\prime}}\delta_{a b}\delta\left(\mathbf{r}-\mathbf{r}^{\prime}\right)\delta\left(\tau-\tau^{\prime}\right). 
\label{eq:eom_1}        
\end{eqnarray}
\end{widetext}
When we follow a similar procedure to write Eq.~(\ref{eq:eom_psi}) for $\psi_{j\sigma}^\dagger(\mathbf{r}, \tau)$, we find another equation of motion as follows
\begin{widetext}
\begin{eqnarray}
        &&{\left[\frac{\partial}{\partial \tau}+\frac{\mathbf{\nabla}^2}{2 m}+\mu\right] F_{j j^{\prime}}^{\dagger ab}\left(\mathbf{r}, \tau; \mathbf{r}^{\prime}, \tau^\prime\right)} 
        -\frac{J}{2}\left\{ F_{j+1, j^{\prime}}^{\dagger ab}\left(\mathbf{r}, \tau; \mathbf{r}^{\prime}, \tau^\prime\right)
        + F_{j-1, j^{\prime}}^{\dagger ab}\left(\mathbf{r}, \tau; \mathbf{r}^{\prime}, \tau^\prime\right)\right\} \nonumber\\
        &&+\lambda \sum_\gamma \left\{ F_{jj}^{\dagger a \gamma}(0+) G_{jj^\prime}^{\gamma b}\left(\mathbf{r}, \tau; \mathbf{r}^{\prime}, \tau^\prime\right) 
        - G_{j j}^{\gamma\gamma}\left(0+\right) F_{j j^{\prime}}^{\dagger a b}\left(\mathbf{r}, \tau; \mathbf{r}^{\prime}, \tau^\prime\right) 
        + G_{jj}^{\gamma a}(0+) F_{jj^\prime}^{\dagger  \gamma b}\left(\mathbf{r}, \tau; \mathbf{r}^{\prime}, \tau^\prime\right) \right\}\nonumber\\
        && -\sum_\gamma \left\{\alpha \left(\left(\sigma_x\right)_{a \gamma} k_y - \left(\sigma_y\right)_{a \gamma} k_x \right) 
        + \beta \left(\left(\sigma_x\right)_{a \gamma} k_x - \left(\sigma_y\right)_{a \gamma} k_y \right) \right\} F^{\dagger\gamma b}_{j j^{\prime}}\left(\mathbf{r}, \tau; \mathbf{r}^{\prime}, \tau^\prime\right) \nonumber\\
        &&=0.
\label{eq:eom_2}  
\end{eqnarray}
\end{widetext}
Eqs.~(\ref{eq:eom_1}) and (\ref{eq:eom_2}) are the equations of motion for the temperature Green's functions (\ref{eq:def_normal_green}) -- (\ref{eq:def_anomalous_green_2}) in the presence of Rashba and Dresselhaus SOIs in layered superconductors.
Here, it is important to emphasize that these equations of motion are written in coordinate space. Similar equations of motion can be found in \cite{Klemm1975} for $\lambda = 0$ and in \cite{EPNakhmedov_1994} and \cite{Bulaevskii1976} for $\lambda \neq 0$. In \cite{Bulaevskii1976}, Fourier transforms were also performed, and analytical expressions for the resulting normal and anomalous Green’s functions were obtained. In Sec.~\ref{sec:3}, we will obtain results similar to those in \cite{Bulaevskii1976} from Eqs.~(\ref{eq:eom_1}) and (\ref{eq:eom_2}).
Additionally, we can approach the problem under consideration using the representation of the spinors, which was done for the Rashba SOI case in \cite{Gorkov2001}. When we do this, we will be working in momentum space, and in the expression for the $\Delta$ gap that we will determine in Sec.~\ref{sec:3}, there will be an additional term $e^{-i\varphi_\mathbf{k}}$ arising from the spinors. For the case we are examining, $\varphi_\mathbf{k} = \arg\left[-\left(\alpha k_y + \beta k_x\right) + i\left(\alpha k_x + \beta k_y\right)\right]$. However, since we will be performing calculations for $|\Delta|$ throughout this study, this additional term will become irrelevant.
It is also important to emphasize that the final gap equations we calculate will be similar to those in \cite{Gorkov2001} (for comparison, see Eq.~(\ref{eq:gap_main}) in Sec.~\ref{sec:4} below and Eq.~(19) in \cite{Gorkov2001}).
Finally, we remark that in the case of $J = \alpha = \beta = 0$, we can find the equations of motion for pure superconductors without SOIs \cite{abrikosov2012} from Eqs.~(\ref{eq:eom_1}) and (\ref{eq:eom_2})
\begin{subequations} 
\begin{eqnarray}
    &{\left[- \frac{\partial}{\partial \tau}+\frac{\mathbf{\nabla}^2}{2 m}+\mu\right] G_0\left(\mathbf{r}, \tau; \mathbf{r}^{\prime}, \tau^\prime\right)} 
    + \Delta_0 F^{\dagger}_0\left(\mathbf{r}, \tau; \mathbf{r}^{\prime}, \tau^\prime\right)\nonumber\\
    & = \delta\left(\mathbf{r}-\mathbf{r}^{\prime}\right)\delta\left(\tau-\tau^{\prime}\right),
\label{eq:normal_green_0}    
\end{eqnarray}
\begin{equation}
    {\left[\frac{\partial}{\partial \tau}+\frac{\mathbf{\nabla}^2}{2 m}+\mu\right] F^{\dagger}_0\left(\mathbf{r}, \tau; \mathbf{r}^{\prime}, \tau^\prime\right)} 
    - \Delta^*_0 G_0\left(\mathbf{r}, \tau; \mathbf{r}^{\prime}, \tau^\prime\right) = 0,
\label{eq:anomalous_green_0}
\end{equation}
\end{subequations}
where 
\begin{equation}
    \Delta_0 = |\lambda| F_0(0+),\quad \Delta^*_0 = |\lambda| F^{\dagger}_0(0+),
\label{eq:lambda0}
\end{equation}
and $\lambda = - |\lambda|$.

\section{\label{sec:3}Green's functions}

Eqs.~(\ref{eq:eom_1}) and (\ref{eq:eom_2}) allow us to find analytical expressions for the temperature Green's functions in layered superconductors in the presence of Rashba and Dresselhaus SOIs. 
To achieve this, we perform the following Fourier transforms in Eqs.~(\ref{eq:eom_1}) and (\ref{eq:eom_2})
\begin{subequations} 
\begin{eqnarray}
    G_{jj^\prime}^{ab} \left(\mathbf{r}, \tau; \mathbf{r}^{\prime}, \tau^\prime\right) 
    &=&  T \sum_n \int_{-\pi}^{\pi} \frac{dk_z}{2\pi} \int_{-\infty}^{\infty} \frac{d^2k}{\left(2\pi\right)^2} \nonumber\\
    && e^{-i\omega_n \left(\tau-\tau^\prime\right)} e^{-i k_z (j-j^\prime)} 
    e^{-i \mathbf{k} (\mathbf{r}-\mathbf{r}^\prime)}\nonumber\\
    &&G^{ab} \left(\mathbf{k},k_z,\omega_n\right),
\label{eq:green1_1}
\end{eqnarray}
\begin{eqnarray}
    F_{jj^\prime}^{\dagger ab} \left(\mathbf{r}, \tau; \mathbf{r}^{\prime}, \tau^\prime\right) 
    &=&  T \sum_n \int_{-\pi}^{\pi} \frac{dk_z}{2\pi} \int_{-\infty}^{\infty} \frac{d^2k}{\left(2\pi\right)^2} \nonumber\\
    && e^{-i\omega_n \left(\tau-\tau^\prime\right)} e^{-i k_z (j-j^\prime)} 
    e^{-i \mathbf{k} (\mathbf{r}-\mathbf{r}^\prime)}\nonumber\\
    &&F^{\dagger ab} \left(\mathbf{k},k_z,\omega_n\right),    
\label{eq:green1_2}
\end{eqnarray}
\end{subequations}
where the Matsubara frequencies 
$\omega_n = (2n+1) \pi  T $ guarantee the proper Fermi statistics. It should be emphasized that, since we set the interlayer distance to 1, $k_z$ in the Fourier transforms of~(\ref{eq:green1_1}) and (\ref{eq:green1_2}) is a dimensionless momentum. However, incorporating the interlayer distance into all the calculations below is straightforward and can be achieved simply by multiplying $k_z$ by this distance.
Straightforward calculations shown in Appendix \ref{app_1} demonstrate that by writing Eqs.~(\ref{eq:green1_1}) and (\ref{eq:green1_2}) in the equations of motion (\ref{eq:eom_1}) and (\ref{eq:eom_2}), we find the following analytical functions for the temperature Green's functions in the presence of Rashba and Dresselhaus SOIs.
\begin{subequations} 
\begin{eqnarray}
    G^{\uparrow\uparrow} (\mathbf{k},k_z,\omega_n) 
    &=& G^{\downarrow\downarrow} (\mathbf{k},k_z,\omega_n)  \nonumber\\
    &=& \frac{\omega^2_n+\xi_{\mathbf{k} k_z}^2}{g},
\label{eq:analytic_normal_green_upup}
\end{eqnarray}
\begin{eqnarray}
    G^{\uparrow\downarrow} (\mathbf{k},k_z,\omega_n) 
    &=& \left(G^{\downarrow\uparrow}(\mathbf{k},k_z,\omega_n) \right)^* \nonumber\\
    &=& - \frac{i \omega_n+\xi_{\mathbf{k} k_z}}{g} \nonumber\\
    && \times \left\{ \alpha(k_y+ik_x) + \beta(k_x+ik_y)\right\},\nonumber\\
\label{eq:analytic_normal_green_updown}
\end{eqnarray}
\begin{eqnarray}
    F^{\dagger\uparrow\downarrow}(\mathbf{k},k_z,\omega_n) 
    &=& - F^{\dagger\downarrow\uparrow}(\mathbf{k},k_z,\omega_n)  \nonumber\\
    &=& \frac{i \omega_n-\xi_{\mathbf{k} k_z}}{g} \Delta^*,
\label{eq:analytic_anomalous_green}
\end{eqnarray}
\end{subequations}
where 
\begin{subequations}
\begin{eqnarray}
g 
&=& \left(i \omega_n-\xi_{\mathbf{k} k_z}\right)\left(\omega^2_n+\xi_{\mathbf{k} k_z}^2+|\Delta|^2\right) +\left(i \omega_n+\xi_{\mathbf{k} k_z}\right) \xi_{SO}^2,\nonumber\\
\label{eq:g}
\end{eqnarray}
\begin{equation}
    \xi_{\mathbf{k} k_z} = \frac{\mathbf{k}^2}{2m} + J \cos k_z - \mu,
\label{eq:xi}
\end{equation}
\begin{equation}
    \xi_{SO} = \pm \sqrt{\left(\alpha^2+\beta^2\right)\left(k_x^2+k_y^2\right)+4 \alpha \beta k_x k_y},
\label{eq:soc}
\end{equation}
\end{subequations}
and
\begin{equation}
    \Delta = |\lambda| F^{\downarrow\uparrow}_{jj}(0+),\quad \Delta^* = |\lambda| F^{\dagger\uparrow\downarrow}_{jj}(0+),
\label{eq:main_delta}    
\end{equation}
with $\lambda = - |\lambda|$. It should be noted that the $\xi_{SO}$ determined by Eq.~(\ref{eq:soc}) are the eigenenergies of the single-particle Hamiltonian $H_0 + H_\text{R} + H_\text{D}$ for a two-dimensional electron gas in the presence of Rashba and Dresselhaus SOIs \cite{Schliemann2003}. We can see from Eqs.~(\ref{eq:analytic_normal_green_upup}) -- (\ref{eq:analytic_anomalous_green}) that one of the main effects of Rashba and Dresselhaus SOIs on the Green's functions is through the function $g$, shown in Eq.~(\ref{eq:g}). Another notable effect is that while the normal Green's function components $G^{\uparrow\downarrow}$ and $G^{\downarrow\uparrow}$ are zero when $\alpha = \beta = 0$, they become non-zero when $\alpha \neq 0$ and/or $\beta \neq 0$. Note also that in the case of $J = \alpha = \beta = 0$, we can find the solutions of Eqs.~(\ref{eq:normal_green_0}) and (\ref{eq:anomalous_green_0}) from Eqs.~(\ref{eq:analytic_normal_green_upup}) -- (\ref{eq:analytic_anomalous_green})
\begin{subequations} 
\begin{eqnarray}
    G_0 (\mathbf{k},\omega_n) 
    &=& - \frac{i \omega_n+\xi_{\mathbf{k}}}{\omega^2_n+\xi_{\mathbf{k}}^2+\Delta_0^2},
\label{eq:G0}    
\end{eqnarray}
\begin{eqnarray}
    F^{\dagger}_0 (\mathbf{k},\omega_n) 
    &=& \frac{\Delta^*_0}{\omega^2_n+\xi_{\mathbf{k}}^2+\Delta_0^2},
\label{eq:F0}    
\end{eqnarray}
\end{subequations}
where $\xi_{\mathbf{k}} = \frac{\mathbf{k}^2}{2m} - \mu$. It is also worth noting that for the case $J \neq 0$ and $\alpha = \beta = 0$, the results are also the same as in Eqs.~(\ref{eq:G0}) and ~(\ref{eq:F0}), with the exception that $\xi_{\mathbf{k}}$ is replaced by $\xi_{\mathbf{k} k_z}$ in Eqs.~(\ref{eq:G0}) and (\ref{eq:F0}).

It is important to emphasize that as seen from Eq.~(\ref{eq:F0}), since the anomalous Green's functions $F_0(\mathbf{k},\omega_n)$ and $F^{\dagger}_0(\mathbf{k},\omega_n)$ are equal, the quantity $\Delta_0$ determined by the equations in Eq.~(\ref{eq:lambda0}) is also real. Therefore, in Eqs.~(\ref{eq:G0}) and (\ref{eq:F0}), it suffices to write the square of $\Delta_0$ rather than its absolute value. However, as seen from Eq.~(\ref{eq:analytic_anomalous_green}), when $\alpha \neq 0$ and/or $\beta \neq 0$, this equality does not hold, and $\Delta$ is not a real function. In other words, the presence of SOIs makes the $\Delta$ gap a complex function.

\section{\label{sec:4}Gap equation}
In this section, we examine the effects of Rashba and Dresselhaus SOIs on the gap equation. To do this, we rewrite Eq.~(\ref{eq:main_delta}) for $\Delta^*$ using Eq.~(\ref{eq:green1_2})
\begin{equation}
    \Delta^*
    =  |\lambda| T \sum_n \int_{-\pi}^{\pi} \frac{dk_z}{2\pi} \int_{-\infty}^{\infty} \frac{d^2k}{\left(2\pi\right)^2} F^{\dagger\uparrow\downarrow} \left(\mathbf{k},k_z,\omega_n\right). 
\label{eq:delta_*}
\end{equation}
By substituting Eq.~(\ref{eq:analytic_anomalous_green}) into Eq.~(\ref{eq:delta_*}), we obtain the following gap equation
\begin{equation}
    1 = |\lambda| T \sum_n \int_{-\pi}^{\pi} \frac{dk_z}{2\pi} \int_{-\infty}^{\infty} \frac{d^2k}{\left(2\pi\right)^2} \frac{i\omega_n-\xi_{\mathbf{k} k_z}}{g}.
\label{eq:delta_*_help}    
\end{equation}
Eq.~(\ref{eq:delta_*_help}) shows that the effects of Rashba and Dresselhaus SOIs on the $\Delta$ gap function determined by Eq.~(\ref{eq:main_delta}) are analogous. If we consider the effect of only one SOI, whether it is Rashba or Dresselhaus, the same result for $\Delta$ is obtained in both cases. In the case of two SOIs, as seen from Eq.~(\ref{eq:soc}), if the values of $\alpha$ and $\beta$ are such that, for example, $\alpha = \alpha_0$ and $\beta = \beta_0$, and then they are swapped, i.e., $\alpha = \beta_0$ and $\beta = \alpha_0$, $\Delta$ remains unchanged.
If we consider this equivalent form of Eq.~(\ref{eq:g})
\begin{eqnarray}
    g 
    &=& \left(i \omega_n-\xi_{\mathbf{k} k_z}\right)\left(\omega^2_n+\xi_{\mathbf{k} k_z}^2+|\Delta|^2+\xi_{SO}^2\right)\nonumber\\
    && \times \left(1-\frac{1}{\left(-i \omega_n+\xi_{\mathbf{k} k_z}\right) \frac{\omega^2_n+\xi_{\mathbf{k} k_z}^2+|\Delta|^2+\xi_{SO}^2}{2 \xi_{SO}^2 \xi_{\mathbf{k} k_z}}}\right),
\label{eq:g_help}
\end{eqnarray}
by writing Eq.~(\ref{eq:g_help}) in Eq.~(\ref{eq:delta_*_help}), we can express the gap equation as follows
\begin{equation}
    1 
    = \sum_{\nu=0}^{\infty} (-1)^\nu I_\nu,
\label{eq:gap_eq_help}
\end{equation}
where
\begin{eqnarray}
    I_\nu
    &=& \frac{|\lambda| T}{2} \sum_n \int_{-\pi}^{\pi} \frac{dk_z}{2\pi} \int_{-\infty}^{\infty} \frac{d^2k}{\left(2\pi\right)^2} 
    \frac{1}{\omega^2_n+\xi_{\mathbf{k} k_z}^2+|\Delta|^2+\xi_{SO}^2}\nonumber\\
    && \times \left( \frac{1}{\left(\left(i \omega_n-\xi_{\mathbf{k} k_z}\right)\frac{\omega^2_n+\xi_{\mathbf{k} k_z}^2+|\Delta|^2+\xi_{SO}^2}{2 \xi_{SO}^2 \xi_{\mathbf{k} k_z}}\right)^\nu} \right.\nonumber\\
    &&+ \left. \frac{1}{\left(\left(-i \omega_n-\xi_{\mathbf{k} k_z}\right)\frac{\omega^2_n+\xi_{\mathbf{k} k_z}^2+|\Delta|^2+\xi_{SO}^2}{2 \xi_{SO}^2 \xi_{\mathbf{k} k_z}}\right)^\nu} \right).
\label{eq:i_nu}    
\end{eqnarray}
It is important to emphasize that when writing Eqs.~(\ref{eq:gap_eq_help}) and (\ref{eq:i_nu}), the inequality
\begin{equation}
    \left|\left(\pm i \omega_n-\xi_{\mathbf{k} k_z}\right) \frac{\omega^2_n+\xi_{\mathbf{k} k_z}^2+|\Delta|^2+\xi_{SO}^2}{2 \xi_{SO}^2 \xi_{\mathbf{k} k_z}}\right| > 1
\label{eq:con_1}
\end{equation}
is assumed to hold for each value of $n, \mathbf{k}$, and $k_z$. To show the validity of this, we rewrite the inequality (\ref{eq:con_1}) as follows
\begin{equation}
     \frac{\omega_n^2+\xi_{\mathbf{k} k_z}^2+|\Delta|^2}{\xi_{SO}^2} > \frac{2}{\sqrt{1+\frac{\omega_n^2}{\xi_{\mathbf{k} k_z}^2}}}-1.
\label{eq:con_2}    
\end{equation}
As seen, the left side of the inequality (\ref{eq:con_2}) is always positive. Therefore, the inequality (\ref{eq:con_2}) is always valid when $\omega_n^2 \geq 3 \xi_{\mathbf{k} k_z}^2$, since in this case, the right side of inequality (\ref{eq:con_2}) is less than or equal to zero. 
For $\omega_n^2 < 3 \xi_{\mathbf{k} k_z}^2$, the inequality~(\ref{eq:con_2}) leads to $c_1 \xi_{\mathbf{k} k_z}^2 + |\Delta|^2 > c_2 \xi_{SO}^2$, where $1 \leq c_1 < 4$ and $0 < c_2 \leq 1$. It should be emphasized that the values $c_1 = c_2 = 1$ occur in the case of $\omega_n = 0$. This scenario holds at $T=0$, where, for 2D systems, the average kinetic energy equals half the Fermi energy. In general, experimental studies on such systems demonstrate that while the Fermi energy is on the order of a few eV, the SOI energy is significantly smaller, typically on the order of a few meV \cite{LaShell1996, PhysRevB.69.241401, Lin2019}. With this, we can conclude that the inequality $\xi_{\mathbf{k} k_z}^2 + |\Delta|^2 > \xi_{SO}^2$ holds in almost all cases at $T=0$.
As $\omega_n$ increases, $c_1$ begins to increase, while $c_2$ starts to decrease. As a result, the left side of the inequality $c_1 \xi_{\mathbf{k} k_z}^2 + |\Delta|^2 > c_2 \xi_{SO}^2$ begins to take consistently larger values, while the right side starts to take consistently smaller values. This also shows that the inequality under consideration becomes even more strongly supported as $\omega_n$ increases.

The inequality (\ref{eq:con_1}), we have just proven, also demonstrates that the integral $I_0$ has the largest integrand among the integrals in (\ref{eq:i_nu}). Our calculations have also shown that the $ I_0 $ integral is significantly larger than the other $ I_\nu $ integrals for $ \nu \geq 1 $. For instance, the computations in Appendix~\ref{app_2} reveal that the contribution of the $ I_1 $ integral to the results provided by the $ I_0 $ integral is exceedingly small. Therefore, we neglect the contribution of the $ I_1 $ integral in our calculations. Additionally, the integrands of the integrals $ I_2 $, $ I_3 $, etc., are even smaller than $ I_1 $, allowing us to neglect their contributions as well. Therefore, in our calculations, we focus on the integral $I_0$, which provides the most significant contribution to the gap equation~(\ref{eq:gap_eq_help}) in the presence of SOIs, and we assume $I_0 = 1$. In other words,
\begin{equation}
    1 = |\lambda| T \sum_n \int_{-\pi}^{\pi} \frac{dk_z}{2\pi} \int_{-\infty}^{\infty} \frac{d^2k}{\left(2\pi\right)^2} 
    \frac{1}{\omega^2_n+\xi_{\mathbf{k} k_z}^2+|\Delta|^2+\xi_{SO}^2}.
\label{eq:gap_main}    
\end{equation}
Note that from Eq.~(\ref{eq:gap_main}), we can also derive the exact gap equation for the case without SOIs \cite{abrikosov2012} by setting $\xi_{SO} = 0$.

Finally, we perform the summation over $n$. We know that a function like $ \sum_{n = - \infty}^{\infty} f(n+1/2) $ can be expressed as the sum of residues of $ \pi \tan\pi z f(z) $ at all the poles of $ f(z) $, where $ n+1/2 = z $ \cite{ARFKEN2013}. The poles of the function $f(z) = 1/(\omega^2_n+\xi_{\mathbf{k} k_z}^2+|\Delta|^2+\xi_{SO}^2)$ are $z_{1,2} = \pm i \sqrt{\xi_{\mathbf{k} k_z}^2 + |\Delta|^2 + \xi_{SO}^2} /2\pi T$. Then, we can write the Eq.~(\ref{eq:gap_main}) as follows
\begin{equation}
    1 = \frac{|\lambda|}{2} \int_{-\pi}^{\pi} \frac{dk_z}{2\pi} \int_{-\infty}^{\infty} \frac{d^2k}{\left(2\pi\right)^2} 
    \frac{\tanh \left(\frac{\sqrt{\xi_{\mathbf{k} k_z}^2 + |\Delta|^2 + \xi_{SO}^2}}{2T}\right)}{\sqrt{\xi_{\mathbf{k} k_z}^2 + |\Delta|^2 + \xi_{SO}^2}}.
\label{eq:gap_main_2}    
\end{equation}

Before proceeding to solve Eq.~(\ref{eq:gap_main_2}), we emphasize that the solutions derived in the following sections are valid for values satisfying $J<\mu$. This is because, as shown in Appendix~\ref{app_3}, for $J<\mu$, the Fermi surface takes finite values for all values of $k_z$. In the special case of $J=\mu$,  the Fermi surface does not form only at $k_z=0$. Therefore, extending our calculations to the $J=\mu$ case is possible by neglecting the contribution that would arise from the $k_z = 0$ condition.

\section{\label{sec:5}Zero temperature results}

As noted earlier, in the cases where $\alpha \neq 0$ and/or $\beta \neq 0$, $\Delta$ is a complex function because $F^{\downarrow\uparrow} \neq F^{\dagger\uparrow\downarrow}$ in Eq.~(\ref{eq:main_delta}). Therefore, the main focus of our calculations will be on how the magnitude of $\Delta$ changes with the presence of SOIs. For this purpose, we first examine how the magnitude $|\Delta|$ changes at zero temperature under the influence of Rashba and Dresselhaus SOIs. Knowing that in this limit $\tanh \left(\frac{\sqrt{\xi_{\mathbf{k} k_z}^2 + |\Delta|^2 + \xi_{SO}^2}}{2T}\right) \rightarrow 1$, if we switch to polar coordinates in $\mathbf{k}$ space, we obtain the following integral from Eq.~(\ref{eq:gap_main_2})
\begin{equation}
    1 = \frac{|\lambda| m}{2} \int_{-\pi}^\pi \frac{dk_z}{2\pi} \int_{-\pi}^{\pi} \frac{d\theta}{2\pi} \int_{-\omega_D}^{\omega_D} \frac{d\epsilon}{2\pi} 
    \frac{1}{\sqrt{\epsilon^2+2\eta_1\epsilon+\eta_2}},
\label{eq:T=0_first}    
\end{equation}
where
\begin{subequations}
\begin{equation}
    \epsilon = \frac{k_F}{m} (k-k_F),
\label{eq:epsilon}    
\end{equation}
\vspace{0em}
\begin{equation}
    \eta_1 = J \cos k_z + m \left(\alpha^2+\beta^2+2\alpha\beta\sin 2\theta\right),
\label{eq:lambda1}
\end{equation}
\begin{equation}
    \eta_2 = |\Delta|^2 + J^2 \cos^2 k_z + 2 m \mu \left(\alpha^2+\beta^2+2\alpha\beta\sin 2\theta\right).
\label{eq:lambda2}
\end{equation}
\end{subequations}
It is important to note here that to avoid divergence in the integral of Eq.~(\ref{eq:gap_main_2}), a cutoff is introduced. This cutoff is based on the condition that in the current model, only electrons within an energy range of 2$\omega_D$ around the Fermi surface are involved in the interaction. If we integrate Eq.~(\ref{eq:T=0_first}) with respect to $\epsilon$ within these cutoff limits, we obtain
\begin{equation}
    1 = - \frac{|\lambda| m}{2(2\pi)} \int_{-\pi}^\pi \frac{dk_z}{2\pi} \int_{-\pi}^{\pi} \frac{d\theta}{2\pi} 
    \ln \left(\frac{\eta_3+\eta_4 \sin 2\theta}{4\omega_D^2}\right),
\label{eq:gap_T=0}
\end{equation}
where
\begin{subequations}
\begin{equation}
    \eta_3 = |\Delta|^2+2m(\alpha^2+\beta^2)(\mu-J\cos k_z),
\label{eq:lambda3}    
\end{equation}
\begin{equation}
    \eta_4 = 4m\alpha\beta(\mu-J\cos k_z).
\label{eq:lambda4}   
\end{equation}
\end{subequations}
Eq.~(\ref{eq:gap_T=0}) shows that at $T=0$, the contribution of $J$ to the gap function occurs only in the presence of a SOI.

By substituting Eq.~(\ref{eq:app_3_main}), calculated in Appendix~\ref{app_4}, into Eq.~(\ref{eq:gap_T=0}) and performing some routine calculations, we find that the gap equation depends solely on the integral with respect to $k_z$
\begin{widetext}
\begin{eqnarray}
    1 =- |\lambda| \rho_0^{2D} \int_{-\pi}^\pi \frac{dk_z}{2\pi}
    \ln \left(\frac{\sqrt{|\Delta|^2+2m(\mu-J\cos k_z)(\alpha+\beta)^2}+\sqrt{|\Delta|^2+2m(\mu-J\cos k_z)(\alpha-\beta)^2}}{4\omega_D}\right),
\label{eq:gap_T=0_main}
\end{eqnarray}
\end{widetext}
where $\rho_0^{2D} = m/2\pi$. 

\subsection{Anisotropic gap function}

Our model is based on intralayer pairing interaction. For such a system, in the absence of SOIs, the gap is isotropic \cite{Gulacsi1988, Schneider1989, Liu1992}. From Eq.~(\ref{eq:gap_T=0_main}), we observe that the presence of Rashba and Dresselhaus SOIs, or even just one of them, makes the gap function anisotropic. This anisotropy is related to the extra $\cos k_z$ factors that the $k_z$ momentum along the $z$ direction contributes to the magnitude of the gap function. 
In this section, the possible values of the magnitude of the gap function at different values of $k_z$ are considered. For this purpose, we define the gap function as follows
\begin{equation}
    \Delta (k_z) = |\lambda| F^{\downarrow\uparrow}_{jj} (0+,k_z), \quad 
    \Delta^* (k_z) = |\lambda|  F^{\dagger\uparrow\downarrow}_{jj} (0+,k_z).
\label{eq:anistropic_delta}
\end{equation}
Here, the function $F^{\dagger\uparrow\downarrow}_{jj} (0+,k_z)$ is determined by Eq.~(\ref{eq:green1_2}), just as in the definition given in Eq.~(\ref{eq:main_delta}). However, the key difference is that integration over $k_z$ is not performed in this case. Similarly, Eq.~(\ref{eq:analytic_anomalous_green}) remains valid for $F^{\dagger\uparrow\downarrow}(\mathbf{k},k_z,\omega_n)$, but with the distinction that the $\Delta$ terms in Eq.~(\ref{eq:analytic_anomalous_green}) are now expressed using $\Delta (k_z)$. 

\begin{figure}
\includegraphics[width=0.47\textwidth]{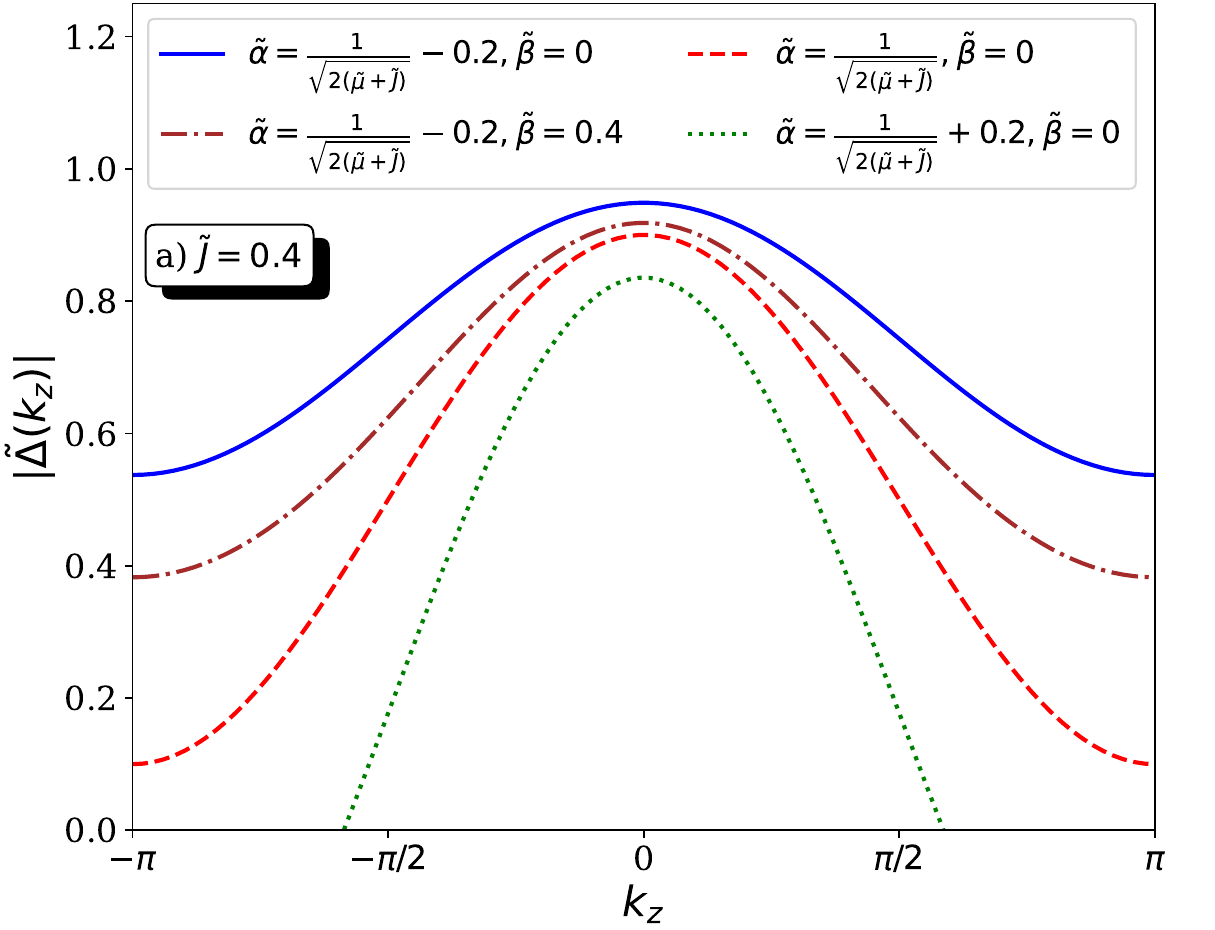}
\hspace{\fill}
\includegraphics[width=0.47\textwidth]{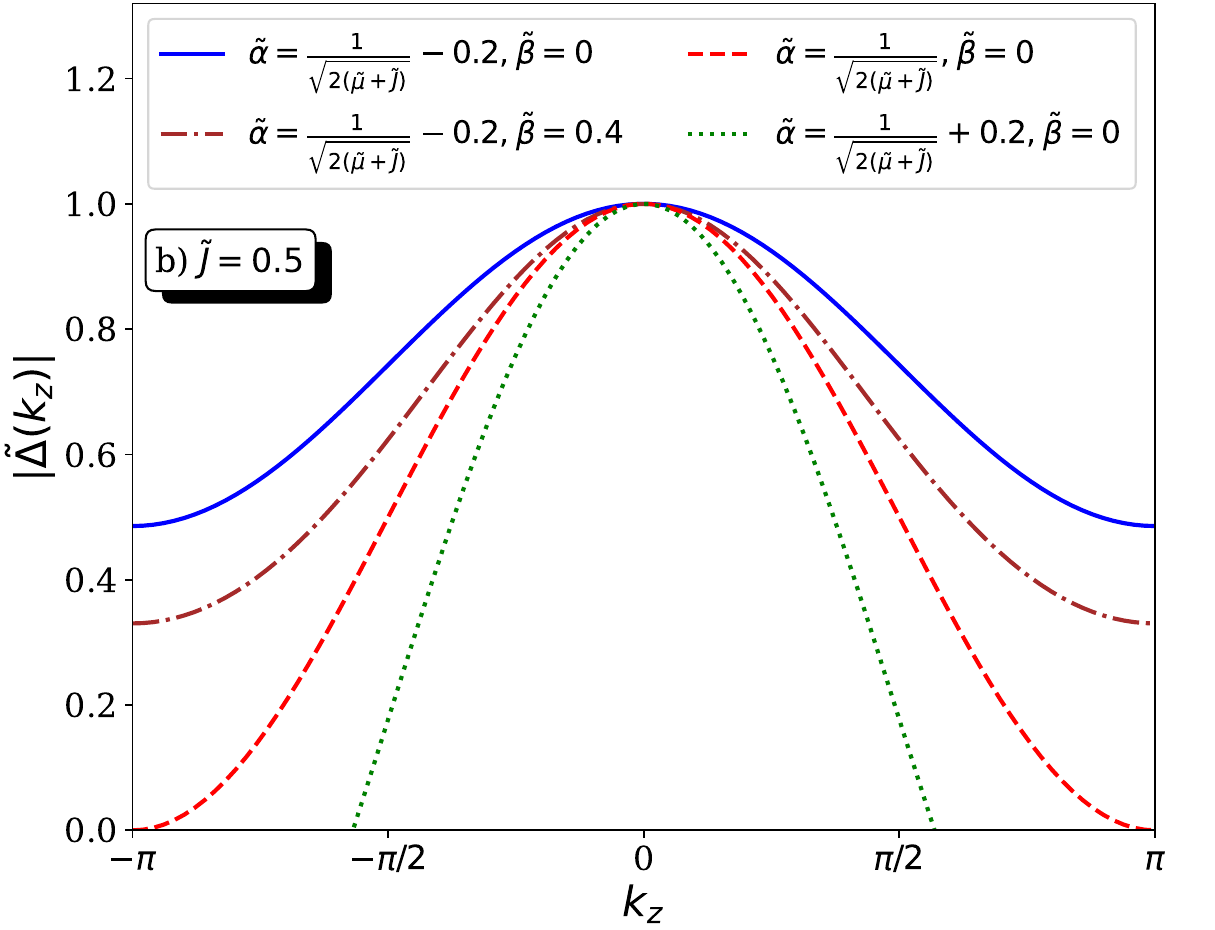}
\hspace{\fill}
\includegraphics[width=0.47\textwidth]{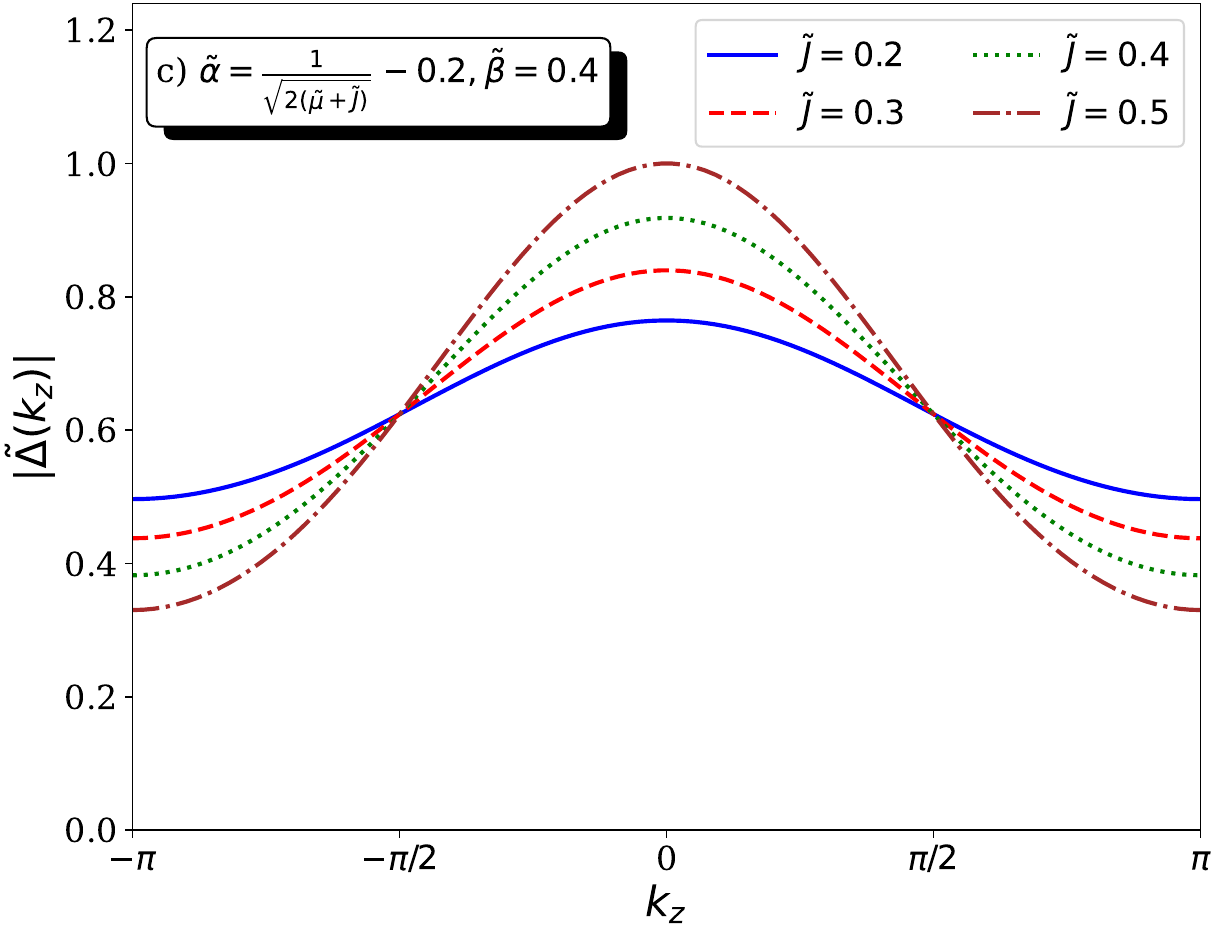}
\caption{Variation of the magnitude of the gap function determined by Eq.~(\ref{eq:anistropic_delta}) with respect to $k_z$ momentum along the $z$ direction. Calculations for all curves are performed with $\tilde{\mu} = 0.5$.}
\label{fig:anistropic}
\end{figure}

By writing the gap equation for the gap function defined in Eq.~(\ref{eq:anistropic_delta}) and carrying out the steps between Eq.~(\ref{eq:delta_*}) and Eq.~(\ref{eq:gap_T=0_main}), we obtain
\begin{widetext}
\begin{equation}
    |\Delta (k_z)|^2 = \Delta_0^2 - 2 m \mu (\alpha^2+\beta^2) + \frac{4 m^2 \mu^2 \alpha^2 \beta^2}{\Delta_0^2} + 2 \left(m (\alpha^2+\beta^2) - \frac{4 m^2 \mu \alpha^2 \beta^2}{\Delta_0^2} \right)J\cos k_z + \frac{4 m^2 \alpha^2 \beta^2}{\Delta_0^2} J^2 \cos^2 k_z,
\label{eq:anistropic}
\end{equation}
\end{widetext}
where $\Delta_0 = 2\omega_De^{-\frac{1}{|\lambda|\rho_0^{2D}}}$. As seen from Eq.~(\ref{eq:anistropic}), when both SOIs are zero, $\Delta = \Delta_0$ and the gap is isotropic. In the presence of one SOI, the anisotropy in $|\Delta (k_z)|^2$ arises from the $\cos k_z$ term, and in the presence of both SOIs, the $\cos^2 k_z$ term also contributes to the anisotropy. This anisotropy is illustrated in Fig.~\ref{fig:anistropic} with the following dimensionless parameters. 
\begin{equation}
    |\tilde{\Delta}| = \frac{\Delta}{\Delta_0}; \tilde{J} = \frac{J}{\Delta_0}; \tilde{\mu} = \frac{\mu}{\Delta_0}; \tilde{\alpha} = \sqrt{\frac{m}{\Delta_0}}\alpha; \tilde{\beta} = \sqrt{\frac{m}{\Delta_0}}\beta.
\label{eq:dimensionless}
\end{equation}
From the comparison of Figs.~\ref{fig:anistropic}a, \ref{fig:anistropic}b, and \ref{fig:anistropic}c we can see that as $J$ increases for fixed values of $\alpha$ and $\beta$, $|\Delta(k_z)|$ increases in the interval $-\pi/2 \leq k_z \leq \pi/2$. The results indicate that, within this interval, electron tunneling between the layers enhances the anisotropic intralayer pairing interaction in the presence of SOI. It is worth noting that, when compared with Fig.~\ref{fig:Fermi_sur}, we can see that for this interval, an increase in $J$ also causes the Fermi surface to form at smaller values of $\mathbf{k}$.
When $k_z = 0$ and $\mu = J$, $|\Delta(k_z)|$ equals $\Delta_0$. This scenario is depicted in Fig.~\ref{fig:anistropic}b, where all curves intersect at the point $|\Delta(k_z)| = \Delta_0$. The reason for this is that, as we see from Fig.~\ref{fig:Fermi_sur}, when $J=\mu$, the Fermi surface does not form only for the case $k_z = 0$. In other words, in this case, the variation of $\alpha$ and $\beta$ has no effect on superconductivity.
In contrast to $J$, larger values of SOIs always suppress $|\Delta(k_z)|$. If we consider $\alpha \neq 0$ and $\beta = 0$, then at $k_z = \pm \pi$, if $\alpha = \Delta_0 / \sqrt{2m(\mu+J)}$, $\Delta(k_z)$ becomes zero for $J=\mu$. This is illustrated by the dashed curve in Fig.~\ref{fig:anistropic}b. For values of $\alpha$ greater than this threshold, $|\Delta(k_z)|$ takes values in a narrower range of $k_z$ within $(- \pi, \pi)$, as shown by the dotted curve in Fig.~\ref{fig:anistropic}. A similar pattern is observed when both $\alpha \neq 0$ and $\beta \neq 0$, with the difference being that terms related to $\cos^2 k_z$ also come into play, resulting in a stronger suppression of $|\Delta(k_z)|$. This last case is illustrated by the dash-dotted curve in Fig.~\ref{fig:anistropic}.

\subsection{Critical SOIs}

\begin{figure}
\includegraphics[width=0.47\textwidth]{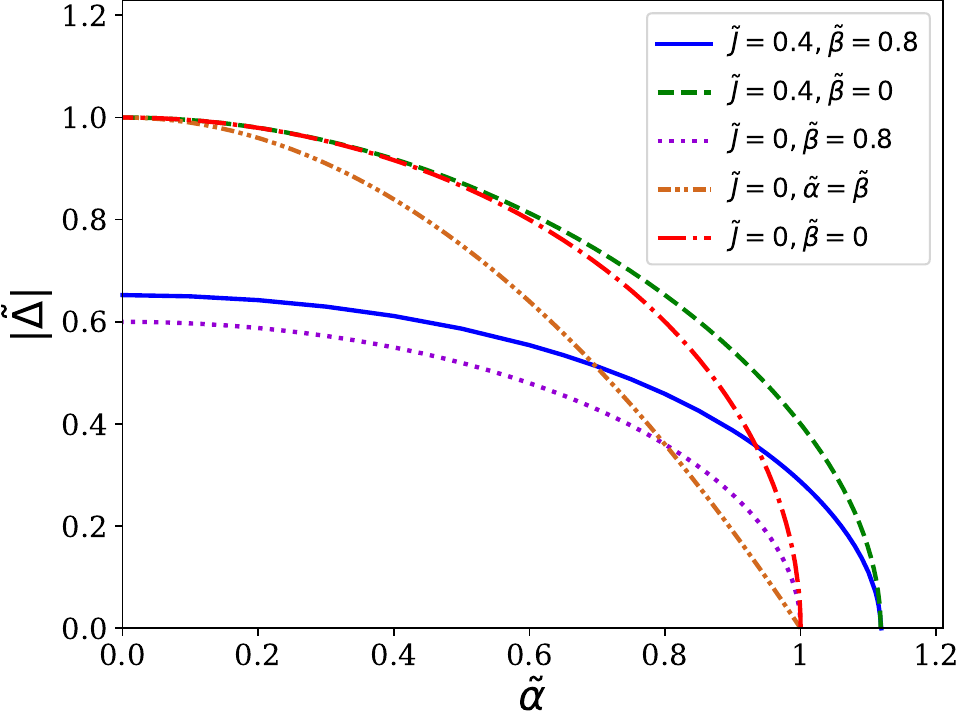}
\caption{Variation of the magnitude of the gap function with the strength of spin-orbit interaction at zero temperature. Calculations for all curves are performed with $\tilde{\mu} = 0.5$.}
\label{fig:T=0}
\end{figure}

In this section, we examine the dependence of the magnitude of the gap function on the strength of SOI. As noted in Sec.~\ref{sec:4}, the effects of Rashba and Dresselhaus SOIs on the gap function are analogous, so we focus on the dependence of $|\Delta|$ on $\alpha$. To this end, we integrate Eq.~(\ref{eq:gap_T=0_main}) with respect to $k_z$. This can be performed numerically for cases where $J \neq 0$, $\alpha \neq 0$, and $\beta \neq 0$. In the case where $J = 0.4 \Delta_0$ and $\beta = 0.8 \sqrt{\Delta_0/m}$, these results are illustrated by the solid curve in Fig.~\ref{fig:T=0}. It is evident from Fig.~\ref{fig:T=0} that SOIs suppresses $|\Delta|$. Moreover, Fig.~\ref{fig:T=0} also indicates that there is a critical value of SOI. If any SOI reaches or exceeds this value, the superconducting phase disappears. By comparing the solid curve with the dashed curve, which represents the case where $ J = 0.4 \Delta_0 $ and $ \beta = 0 $, we observe that the critical value of one SOI is unaffected by the presence of the other SOI. Therefore, the critical value for $\alpha$ calculated in Eq.~(\ref{eq:gap_T=0_main}) for the case $\beta = 0$ will also be valid for $\beta \neq 0$. To calculate this value, assume $\beta = 0$ in the integral in Eq.~(\ref{eq:gap_T=0_main})
\begin{equation}
    1 
    =- \frac{|\lambda| \rho_0^{2D}}{2} \int_{-\pi}^\pi \frac{dk_z}{2\pi}
    \ln \left(\frac{|\Delta|^2+2m(\mu-J\cos k_z)\alpha^2}{4\omega_D^2}\right).
\label{eq:gap_T=0_alpha}
\end{equation}
To find the integral with respect to $k_z$ in Eq.~(\ref{eq:gap_T=0_alpha}), we can follow a method very similar to the one shown in Appendix~\ref{app_4}. As a result, we obtain the following expression
\begin{eqnarray}
    &1 
    =-|\lambda| \rho_0^{2D}\nonumber\\
    & \times\ln \left(\frac{\sqrt{|\Delta|^2+2m(\mu+J)\alpha^2}+\sqrt{|\Delta|^2+2m(\mu-J)\alpha^2}}{4\omega_D}\right).
\label{eq:gap_T=0_alpha1}
\end{eqnarray}
From this, we find that the square of the magnitude of the gap function is
\begin{equation}
    |\Delta|^2 = \Delta_0^2-2m\mu\alpha^2+\frac{\left(mJ\alpha^2\right)^2}{\Delta_0^2}.
\label{eq:gap_T=0_alpha_main}
\end{equation}
At zero temperature, the critical value of $\alpha$ at which the superconducting phase vanishes in the system described by the Hamiltonian (\ref{eq:main_H}) can be determined from Eq.~(\ref{eq:gap_T=0_alpha_main}) as follows
\begin{equation}
    \alpha_C^J = \frac{\Delta_0}{J}\sqrt{\frac{\mu}{m}\left(1-\sqrt{1-\frac{J^2}{\mu^2}}\right)}.
\label{eq:cri_J}
\end{equation}
As shown in Eq.~(\ref{eq:cri_J}), the critical value of the SOI depends on $J$. Generally, as $J$ increases, this critical value also increases. This is because, as noted for Fig.~\ref{fig:anistropic}, an increase in $J$ leads to a rise in $|\Delta|$, necessitating a stronger SOI to nullify it. This behavior is also evident from the comparison of $J=0$ and $J \neq 0$ cases, where $J=0$ represents the scenario of pure $2D$ superconductors in the presence of Rashba and Dresselhaus SOIs. For $J=0$, $\alpha \neq 0$, and $\beta \neq 0$, the solution to Eq.~(\ref{eq:gap_T=0_main}) can be obtained by setting $J=0$ in Eq.~(\ref{eq:anistropic})
\begin{equation}
    |\Delta|^2 = \Delta_0^2 -2m\mu(\alpha^2+\beta^2) + \frac{4m^2\mu^2\alpha^2\beta^2}{\Delta_0^2 }.
 \label{eq:J=0_main}   
\end{equation}
Note that Eq.~(\ref{eq:J=0_main}) simplifies to the following form when $\alpha = \beta$
\begin{equation}
    |\Delta| = \Delta_0 - \frac{2m\mu\alpha^2}{\Delta_0}.
\label{eq:J=0_alpha_beta}
\end{equation}
From Eq.~(\ref{eq:J=0_main}), we can determine the critical SOI value for pure $2D$ superconductors
\begin{equation}
    \alpha_C^{2D} = \frac{\Delta_0}{\sqrt{2m\mu}}.
\label{eq:cri_2d}
\end{equation}
As seen in Fig.~\ref{fig:T=0}, the critical SOI value for the $J=0$ curves is smaller compared to the case where $J \neq 0$. Generally, for the same value of $\mu$
\begin{equation}
    \alpha_C^J > \alpha_C^{2D}.
\end{equation}
It should be noted that, in Fig.~\ref{fig:T=0}, with $\tilde{\mu} = 0.5$, the critical SOI value $\tilde{\alpha}_C^{2D}$ is 1.

\section{\label{sec:6}Finite temperature results}

In this section, we examine how the magnitude of $|\Delta|$ and the critical temperature $T_C$ change in the presence of Rashba and Dresselhaus SOIs at finite temperatures. To do this, we first express Eq.~(\ref{eq:gap_main_2}) at finite temperatures by converting to polar coordinates in $\mathbf{k}$-space. We then obtain the following equation
\begin{equation}
    1 
    = \frac{|\lambda| \rho_0^{2D}}{2} \int_{-\pi}^\pi \frac{dk_z}{2\pi} \int_{-\pi}^{\pi} \frac{d\theta}{2\pi} \int_{-\omega_D}^{\omega_D} d\epsilon 
    \frac{\tanh\left(\frac{\sqrt{\epsilon^2+2\eta_1\epsilon+\eta_2}}{2T}\right)}{\sqrt{\epsilon^2+2\eta_1\epsilon+\eta_2}}.
\label{eq:T_first}
\end{equation}
By writing  
\begin{equation}
     \tanh\left(\frac{\sqrt{\epsilon^2+2\eta_1\epsilon+\eta_2}}{2T}\right)
     = 1- \frac{2}{1+e^{\frac{\sqrt{\epsilon^2+2\eta_1\epsilon+\eta_2}}{T}}},
\label{eq:tanh}
\end{equation}
we can see that if we study the temperature dependence of  $\Delta(T)$ , then we have to keep at  $T \ll T_C$  the approximate $ e^{-\frac{\sqrt{\epsilon^2+2\eta_1\epsilon+\eta_2}}{T}}$  dependence which is responsible for the  $T$ -dependence of  $|\Delta|$. On the other hand, if we seek an equation for  $T_C$, we neglect this term. Below, we examine these two cases separately.

\subsection{Critical temperature}

The critical temperature can be determined using Eq.~(\ref{eq:T_first}), where the only difference is that we assume $\Delta = 0$ in the expression for $\eta_2$. By performing tedious but straightforward calculations, we obtain the following equation for the critical temperature
\begin{eqnarray}
    1 
    &=& \frac{|\lambda| \rho_0^{2D}}{2} \int_{-\pi}^\pi \frac{dk_z}{2\pi} \int_{-\pi}^{\pi} \frac{d\theta}{2\pi} \left( 2 \ln\left(\frac{2\omega_D e^\gamma}{\pi T_C}\right) -\sqrt{\tilde{\eta}_2-\eta_1^2}\right),\nonumber\\
\label{eq:T_C}
\end{eqnarray}
where $\tilde{\eta}_2 = J^2 \cos^2 k_z + 2 m \mu \left(\alpha^2+\beta^2+2\alpha\beta\sin 2\theta\right)$, and $\gamma$ is the Euler-Mascheroni constant. 
As seen from Eq.~(\ref{eq:T_C}), the contribution of $J$ to $T_C$ is only possible in the presence of SOIs.
By numerically solving Eq.~(\ref{eq:T_C}), we can determine the critical temperature for layered superconductors in the presence of Rashba and Dresselhaus SOIs. These results are presented in Table~\ref{tb:T_C}. As can be seen from this table, the influence of Rashba and Dresselhaus SOIs on the BCS equation is substantial. Generally, SOIs reduce the ratio $T_C/\Delta_0$. The presence of a single SOI causes the most significant suppression, while the presence of a second SOI also leads to suppression, but to a lesser extent. Large values of $J$ slightly increase the $T_C/\Delta_0$ ratio.
\begin{table}
\tabcolsep=0.2cm
\def\arraystretch{1.4}
\caption{BCS equation for layered superconductors in the presence of Rashba and Dresselhaus SOIs. In these calculations, $\tilde{\mu}=0.5$ has been chosen.} \label{tb:T_C}
\begin{tabular}{cccccc}
\hline \hline  $\tilde{\alpha}$ & $\tilde{\beta}$ & $\tilde{J}$ & $T_C / \Delta_0$ \\
\hline  
0 & 0 & 0 & 0.567\\
0.2 & 0 & 0 & 0.514\\
0.2 & 0.2 & 0 & 0.502\\
0.2 & 0.2 & 0.2 & 0.503\\
0.4 & 0 & 0 & 0.472\\
0.4 & 0.2 & 0 & 0.471\\
0.4 & 0.2 & 0.2 & 0.472\\
0.4 & 0.4 & 0 & 0.468\\
0.4 & 0.4 & 0.2 & 0.469\\
0.4 & 0.4 & 0.4 & 0.471\\
\hline \hline
\end{tabular}
\end{table}

\subsection{The gap function at temperatures $T \ll T_C$}

By substituting Eq.~(\ref{eq:tanh}) into Eq.~(\ref{eq:T_first}), we obtain two integrals. The first integral is identical to the one on the right-hand side of Eq.~(\ref{eq:T=0_first}) and, as a result, is equal to the right-hand side of Eq.~(\ref{eq:gap_T=0_main}), with the only difference being that \(\Delta\) is now a function of \(T\). The second integral is given by the following expression
\begin{eqnarray}
    I(T) 
    &=& - |\lambda| \rho_0^{2D} \int_{-\pi}^\pi \frac{dk_z}{2\pi} \int_{-\pi}^{\pi} \frac{d\theta}{2\pi} \int_{-\omega_D}^{\omega_D} d\epsilon\nonumber\\
    &&
    \times\frac{\sum_{n=1}^\infty (-1)^{n+1} e^{-n\frac{\sqrt{\epsilon^2+2\eta_1\epsilon+\eta_2}}{T}}}{\sqrt{\epsilon^2+2\eta_1\epsilon+\eta_2}}.
\label{eq:part_T}
\end{eqnarray}
where we used the following expansion for the small exponential term
\begin{equation}
    \left(1+e^{x}\right)^{-1} = \sum_{n=1}^\infty (-1)^{n+1} e^{-nx}.
\end{equation}
From the integral representation of the modified Bessel function of the second kind \cite{ARFKEN2013} we can write
\begin{equation}
    K_0 \left(\frac{n\sqrt{\eta_2-\eta_1^2}}{T}\right) = \frac{1}{2} \int_{-\infty}^\infty d\epsilon \frac{e^{-\frac{n}{2}\sqrt{\epsilon^2+2\eta_1\epsilon+\eta_2}}}{\sqrt{\epsilon^2+2\eta_1\epsilon+\eta_2}}.
\label{eq:bessel_1}
\end{equation}
On the other hand, we can represent the modified Bessel function of the second kind using the following series
\begin{equation}
    K_0(x) = \sqrt{\frac{\pi}{2x}} e^{-x} \sum_{k=0}^{n-1} \frac{1}{(2x)^k} \frac{\Gamma(\frac{1}{2}+k)}{k!\Gamma(\frac{1}{2}-k)}.
\label{eq:Bes2}
\end{equation}
Since $\frac{n\sqrt{\eta_2-\eta_1^2}}{T} \gg 1$ in the case where $T \ll T_C$, we can approximate the series expansion given by Eq.~(\ref{eq:Bes2}) using only the first two terms. Then
\begin{eqnarray}
    \sum_{n=1}^\infty (-1)^{n+1} K_0 \left(\frac{n\sqrt{\eta_2-\eta_1^2}}{T}\right) 
    &\approx& \sqrt{\frac{\pi T}{2\sqrt{\eta_2-\eta_1^2}}} e^{-\frac{\sqrt{\eta_2-\eta_1^2}}{T}}\nonumber\\
    &&\times \left(1-\frac{T}{8\sqrt{\eta_2-\eta_1^2}}\right).\nonumber\\
\label{eq:bessel_2}
\end{eqnarray}
Thus, by approximating $\pm \omega_D \rightarrow \pm \infty$ and substituting Eq.~(\ref{eq:bessel_2}) in Eq.~(\ref{eq:part_T}), we can obtain Eq.~(\ref{eq:T_first}) as follows
\begin{eqnarray}
    1 
    &=& -\frac{|\lambda| \rho_0^{2D}}{2} \int_{-\pi}^\pi \frac{dk_z}{2\pi} \int_{-\pi}^{\pi} \frac{d\theta}{2\pi} \left\{\ln \left(\frac{\eta_3+\eta_4 \sin 2\theta}{4\omega_D^2}\right)\right.\nonumber\\
    &&\left.+4\sqrt{\frac{\pi T}{2\sqrt{\eta_2 - \eta_1^2}}} e^{-\frac{\sqrt{\eta_2 - \eta_1^2}}{T}}\left(1-\frac{T}{8\sqrt{\eta_2 - \eta_1^2}}\right)\right\}.\nonumber\\
\label{eq:geT_main}
\end{eqnarray}
Eq.~(\ref{eq:geT_main}) represents the gap equation for layered superconductors in the presence of Rashba and Dresselhaus SOIs at finite temperatures where $T \ll T_C$. One significant conclusion from this equation is that, similar to zero temperature, the effect of tunneling of electrons between layers on the intralayer gap is significant in the presence of SOIs at temperatures below the critical temperature. In other words, the anisotropy of the gap function at finite temperatures is also due to the presence of SOIs. 

\begin{figure}
\includegraphics[width=0.47\textwidth]{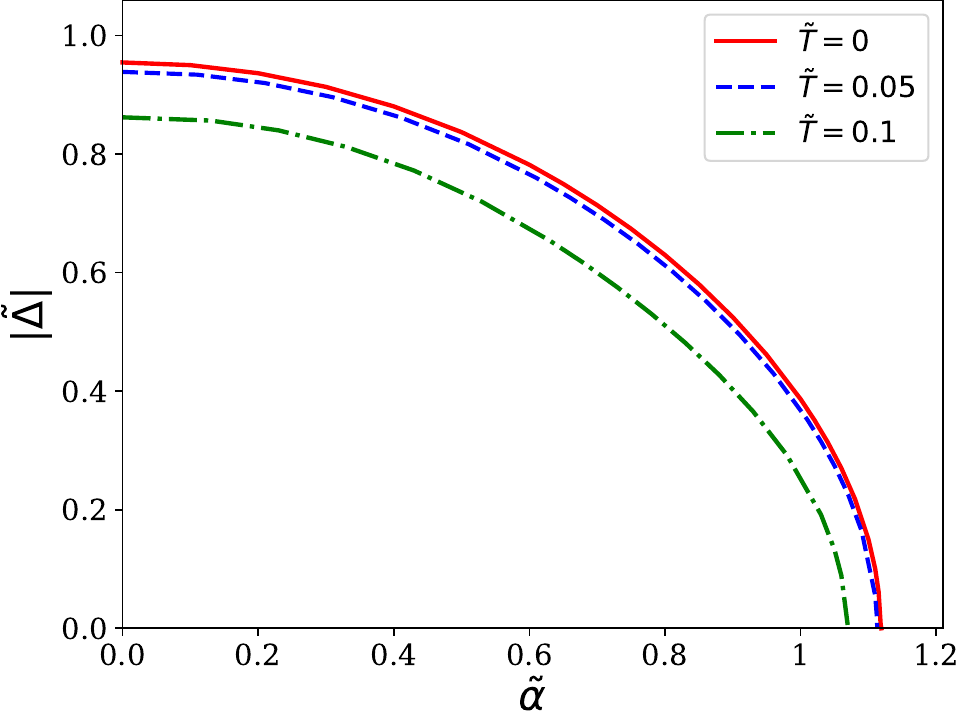}
\caption{Variation of the magnitude of the gap function with the strength of spin-orbit interaction at temperatures $T \ll T_C$. Calculations for all curves are performed with $\tilde{J}=0.4, \tilde{\beta}=0.3$, and $\tilde{\mu} = 0.5$.}
\label{fig:TlessTc}
\end{figure}

Eq.~(\ref{eq:geT_main}) can be solved numerically. Fig.~\ref{fig:TlessTc} illustrates the dependence of $|\Delta|$ on $\alpha$ at temperatures $T = 0$, $0.05 \Delta_0$, and $0.1 \Delta_0$ with $J = 0.4 \Delta_0$ and $\beta = 0.3 \sqrt{\Delta_0/m}$. Fig.~\ref{fig:TlessTc} shows that temperature, in conjunction with SOIs, suppresses $|\Delta|$. Another result indicated by this figure is that the critical SOI value decreases as the temperature increases.

\section{\label{sec:levelc}Conclusions}

In this study we have extended the theory of layered superconductors with intralayer pairing interaction to include Rashba and Dresselhaus SOIs. One of the key results we obtained is that the presence of SOIs renders the gap function anisotropic in these systems. This effect is demonstrated in Eq.~(\ref{eq:anistropic}) for zero temperature and in Eq.~(\ref{eq:geT_main}) for finite temperatures. Generally, our work serves as a foundational theory for experimental studies such as \cite{Yoshizawa2021, Uchihashi2021, PhysRevB.103.245113, SAKAMOTO2022100665, yaji2010large, PhysRevLett.115.147003}. Moreover, the anisotropy introduced by SOIs could be crucial in understanding superconductivity and its anisotropic behavior across various systems. For instance, anisotropic 2D superconductivity has been observed in KTaO$_3$(111) interfaces, with critical temperatures dependent on orientation and sensitivity to bias direction, though the underlying mechanisms are not yet fully understood \cite{Changjiang2021, Ethan2023}. Notably, Rashba-type spin splitting has been reported for the KTaO$_3$(111) surface \cite{Bruno2019}. However, this work has not explained the highly anisotropic nature of superconductivity. Our study, in contrast, demonstrates that additional anisotropy may exist in superconductors as a result of Rashba-type spin splitting.

Another important result of our study is the existence of a critical value for the SOIs, where, when one of the SOIs reaches this value, the superconducting phase in the system ceases. We calculated these values analytically for zero temperature, as provided in Eq.~(\ref{eq:cri_J}). For finite temperatures, the critical values are determined by the numerical solution of Eq.~(\ref{eq:geT_main}) (see Fig.~\ref{fig:TlessTc}). This result indicates that, similar to the Pauli limit, the model exhibits a Rashba and/or Dresselhaus limit in which the paired states will be distrupted. Therefore, in experimental studies, it is necessary to consider that the superconducting phase could be terminated due to the effects of the SOIs.

\begin{acknowledgments}
BT is supported by the Turkish Academy of Sciences (TUBA).
\end{acknowledgments}

\appendix

\section{\label{app_1}Components of the Green’s function in the presence of SOIs}

If we substitute Eqs.~(\ref{eq:green1_1}) and (\ref{eq:green1_2}) into the equations of motion (\ref{eq:eom_1}) and (\ref{eq:eom_2}), we find the following equations
\begin{widetext}
\begin{subequations}
\begin{eqnarray}
    &\left(i \omega_n - \xi_{\mathbf{k} k_z} - \lambda G^{\downarrow\downarrow}(0+)\right) G^{\uparrow\uparrow} (\mathbf{k},k_z,\omega_n) 
    + \left( \alpha(k_y+ik_x)+\beta(k_x+ik_y)+\lambda G^{\uparrow\downarrow}(0+)\right) G^{\downarrow\uparrow} (\mathbf{k},k_z,\omega_n)\nonumber\\
    &- \lambda F^{\uparrow\downarrow}(0+) F^{\dagger\downarrow\uparrow} (\mathbf{k},k_z,\omega_n) = 1,
\label{eq:app1_1}
\end{eqnarray}
\begin{eqnarray}
    \left(i \omega_n - \xi_{\mathbf{k} k_z} - \lambda G^{\downarrow\downarrow}(0+)\right) G^{\uparrow\downarrow} (\mathbf{k},k_z,\omega_n) 
    + \left( \alpha(k_y+ik_x)+\beta(k_x+ik_y)+\lambda G^{\uparrow\downarrow}(0+)\right) G^{\downarrow\downarrow} (\mathbf{k},k_z,\omega_n)\nonumber = 0,\nonumber\\
\label{eq:app1_2}
\end{eqnarray}
\begin{eqnarray}
    \left(i \omega_n - \xi_{\mathbf{k} k_z} - \lambda G^{\uparrow\uparrow}(0+)\right) G^{\downarrow\uparrow} (\mathbf{k},k_z,\omega_n) 
    + \left( \alpha(k_y-ik_x)+\beta(k_x-ik_y)+\lambda G^{\downarrow\uparrow}(0+)\right) G^{\uparrow\uparrow} (\mathbf{k},k_z,\omega_n)\nonumber = 0,\nonumber\\
\label{eq:app1_3}
\end{eqnarray}
\begin{eqnarray}
    &\left(i \omega_n - \xi_{\mathbf{k} k_z} - \lambda G^{\uparrow\uparrow}(0+)\right) G^{\downarrow\downarrow} (\mathbf{k},k_z,\omega_n) 
    + \left( \alpha(k_y-ik_x)+\beta(k_x-ik_y)+\lambda G^{\downarrow\uparrow}(0+)\right) G^{\uparrow\downarrow} (\mathbf{k},k_z,\omega_n)\nonumber\\
    &- \lambda F^{\downarrow\uparrow}(0+) F^{\dagger\uparrow\downarrow} (\mathbf{k},k_z,\omega_n) = 1,
\label{eq:app1_4}
\end{eqnarray}
\begin{eqnarray}
    \left(-i \omega_n - \xi_{\mathbf{k} k_z} - \lambda G^{\downarrow\downarrow}(0+)\right) F^{\dagger\uparrow\downarrow} (\mathbf{k},k_z,\omega_n) 
    + \lambda F^{\dagger\uparrow\downarrow}(0+) G^{\downarrow\downarrow} (\mathbf{k},k_z,\omega_n)\nonumber = 0,\nonumber\\
\label{eq:app1_5}
\end{eqnarray}
\begin{eqnarray}
    \left(-i \omega_n - \xi_{\mathbf{k} k_z} - \lambda G^{\uparrow\uparrow}(0+)\right) F^{\dagger\downarrow\uparrow} (\mathbf{k},k_z,\omega_n) 
    + \lambda F^{\dagger\downarrow\uparrow}(0+) G^{\uparrow\uparrow} (\mathbf{k},k_z,\omega_n)\nonumber = 0,\nonumber\\
\label{eq:app1_6}
\end{eqnarray}
\end{subequations}
\end{widetext}
where $\xi_{\mathbf{k} k_z}$ is determined by Eq.~(\ref{eq:xi}). We note that when evaluating Eqs.~(\ref{eq:app1_5}) and (\ref{eq:app1_6}), we take into account 
\begin{eqnarray}
    &&\sum_\gamma \left\{\alpha \left(\left(\sigma_x\right)_{a \gamma} k_y - \left(\sigma_y\right)_{a \gamma} k_x \right) \right.\nonumber\\
    &&\left.+ \beta \left(\left(\sigma_x\right)_{a \gamma} k_x - \left(\sigma_y\right)_{a \gamma} k_y \right) \right\} F^{\dagger\gamma b}_{j j^{\prime}}\left(\mathbf{r}, \tau; \mathbf{r}^{\prime}, \tau^\prime\right) = 0.\nonumber\\
\label{eq:anom_green_spin}
\end{eqnarray}
in Eq.~(\ref{eq:eom_2}). To understand the reason behind Eq.~(\ref{eq:anom_green_spin}), we begin by assuming spin state $a$ is up ($\uparrow$). Consequently, this leads to spin state $b$ being down ($\downarrow$) in Eq.~(\ref{eq:eom_2}). Therefore, for the left-hand side of Eq.~(\ref{eq:anom_green_spin}) to be non-zero, $\gamma$ must be $\uparrow$. However, both $\sigma_x$ and $\sigma_y$ matrices have zero elements for $\uparrow \uparrow$, indicating the validity of Eq.~(\ref{eq:anom_green_spin}). A similar situation can be observed for $a = \downarrow$ and $b = \uparrow$.

To find the analytical expressions for the Green's functions from Eqs.~(\ref{eq:app1_1}) -- (\ref{eq:app1_6}), we first determine the functions $G^{\downarrow\uparrow} (\mathbf{k},k_z,\omega_n)$ and $F^{\dagger\downarrow\uparrow} (\mathbf{k},k_z,\omega_n)$ from Eqs.~(\ref{eq:app1_3}) and (\ref{eq:app1_6}). Subsequently, by writing them in Eq.~(\ref{eq:app1_1}), we can derive the following equation for $G^{\uparrow\uparrow} (\mathbf{k},k_z,\omega_n)$
\begin{eqnarray}
    &\left(i\omega_n - \xi_{\mathbf{k} k_z} -\frac{(\alpha^2+\beta^2)(k_x^2+k_y^2)+4\alpha\beta k_xk_y}{i\omega_n-\xi_{\mathbf{k} k_z}} +\frac{|\Delta|^2}{-i\omega_n-\xi_{\mathbf{k} k_z}}\right) \nonumber\\
    &\times G^{\uparrow\uparrow}(\mathbf{k},k_z,\omega_n) = 1,
\label{eq:app13}
\end{eqnarray}
where $|\Delta|^2$ is determined by Eq.~(\ref{eq:main_delta}). Similarly, by determining the functions $G^{\uparrow\downarrow} (\mathbf{k},k_z,\omega_n)$ and $F^{\dagger\uparrow\downarrow} (\mathbf{k},k_z,\omega_n)$ from Eqs.~(\ref{eq:app1_2}) and (\ref{eq:app1_5}), and then writing them in Eq.~(\ref{eq:app1_4}), we obtain the same equation for $G^{\downarrow\downarrow} (\mathbf{k},k_z,\omega_n)$ as in Eq.~(\ref{eq:app13}). From this, we establish that $G^{\uparrow\uparrow} (\mathbf{k},k_z,\omega_n) = G^{\downarrow\downarrow} (\mathbf{k},k_z,\omega_n)$, and from Eq.~(\ref{eq:app13}), we find Eq.~(\ref{eq:analytic_normal_green_upup}). Here, note that in Eq.~(\ref{eq:app13}), we do not account for the contributions of the $\lambda G^{\uparrow\uparrow, \downarrow\downarrow}(0+)$ terms, as done in the theory of pure superconductors without SOIs \cite{abrikosov2012}. This is because these terms only add to the chemical potential and hence are of no interest. Additionally, in deriving Eq.~(\ref{eq:app13}), we also neglect the contributions of $\lambda G^{\uparrow\downarrow, \downarrow\uparrow}(0+)$ terms because these contributions are very small, on the order of $\lambda / \mu$. Thus, by obtaining the analytical expressions represented by Eq.~(\ref{eq:analytic_normal_green_upup}) for $G^{\uparrow\uparrow, \downarrow\downarrow} (\mathbf{k},k_z,\omega_n)$, we find Eq.~(\ref{eq:analytic_normal_green_updown}) for the off-diagonal elements of the normal Green’s function from Eqs.~(\ref{eq:app1_2}) and (\ref{eq:app1_3}), and Eq.~(\ref{eq:analytic_anomalous_green}) for the anomalous Green's functions from Eqs.~(\ref{eq:app1_5}) and (\ref{eq:app1_6}).

\section{\label{app_2} Contribution of Integral $I_1$}

In this appendix, we examine the contribution of the $I_1$ integral to $\Delta$. In other words, we assume 
the gap equation to be of the form
\begin{equation}
    1 = I_0 - I_1,
\label{eq:app2_1}
\end{equation}
where the calculations related to $I_0$ have been carried out in Secs.~\ref{sec:5} and \ref{sec:6}, and 
\begin{eqnarray}
    I_1
    &=&- |\lambda| T \sum_n \int_{-\pi}^{\pi} \frac{dk_z}{2\pi} \int_{-\infty}^{\infty} \frac{d^2k}{\left(2\pi\right)^2} 
    \frac{1}{\omega^2_n+\xi_{\mathbf{k} k_z}^2+|\Delta|^2+\xi_{SO}^2}\nonumber\\
    && \times \frac{1}{\left(\omega_n^2+\xi_{\mathbf{k} k_z}^2\right)\frac{\omega^2_n+\xi_{\mathbf{k} k_z}^2+|\Delta|^2+\xi_{SO}^2}{2 \xi_{SO}^2 \xi_{\mathbf{k} k_z}^2}}.
\label{eq:i_1}    
\end{eqnarray}
In order to begin, we perform the summation over $n$ for the integral $I_1$. Since the poles of the integrand are $z_{1,2} = \pm i \sqrt{\xi_{\mathbf{k} k_z}^2 + |\Delta|^2 + \xi_{SO}^2} /2\pi T$ and $z_{3,4} = \pm i \frac{\xi_{\mathbf{k} k_z}}{2\pi T}$, where $z = n + 1/2$, we find that
\begin{widetext}
    \begin{eqnarray}
        I_1
    &=& \frac{|\lambda|}{2} \int_{-\pi}^{\pi} \frac{dk_z}{2\pi} \int_{-\infty}^{\infty} \frac{d^2k}{\left(2\pi\right)^2} \frac{\xi_{SO}^2 \xi_{\mathbf{k} k_z}}{(2\pi T)^4} 
    \left[\frac{(2\pi T)^4 \tanh\left(\frac{\xi_{\mathbf{k} k_z}}{2T}\right)}{(|\Delta|^2+\xi_{SO}^2)^2} 
    - i\left(\frac{1}{(z_1-z_3)^2} - \frac{1}{(z_1-z_4)^2} +\frac{2}{(z_1-z_2)(z_1-z_3)} \right.\right.\nonumber\\
    && \left. - \frac{2}{(z_1-z_2)(z_1-z_4)}
    \right) \frac{\tan(\pi z_1)}{(z_1-z_2)^2} 
    + 2 i \left(\frac{1}{z_2-z_3}-\frac{1}{z_2-z_4}\right) \frac{\tan(\pi z_2)}{(z_1-z_2)^3} 
    + i\pi \left( \frac{1}{(z_1-z_3)(z_2-z_3)^2} - \frac{1}{(z_1-z_4)(z_2-z_4)^2}\right) 
    \nonumber\\
    &&\left.\times \frac{\tan(\pi z_2)}{(z_1-z_2)^4 \cos^2(\pi z_1)} +i\pi \left( \frac{1}{z_2-z_3} - \frac{1}{z_2-z_4} \right)\frac{1}{(z_1-z_2)^2 \cos^2(\pi z_2)} \right. \Bigg].
    \label{eq:app2_3}
    \end{eqnarray}
\end{widetext}
In Eq.~(\ref{eq:app2_3}), we can see how much more complicated the $ I_1 $ integral is compared to $ I_0 $. Note that the $ I_0 $ integral is computed as shown in Eq.~(\ref{eq:gap_main_2}) using similar steps.

\begin{figure}[H]
\includegraphics[width=0.47\textwidth]{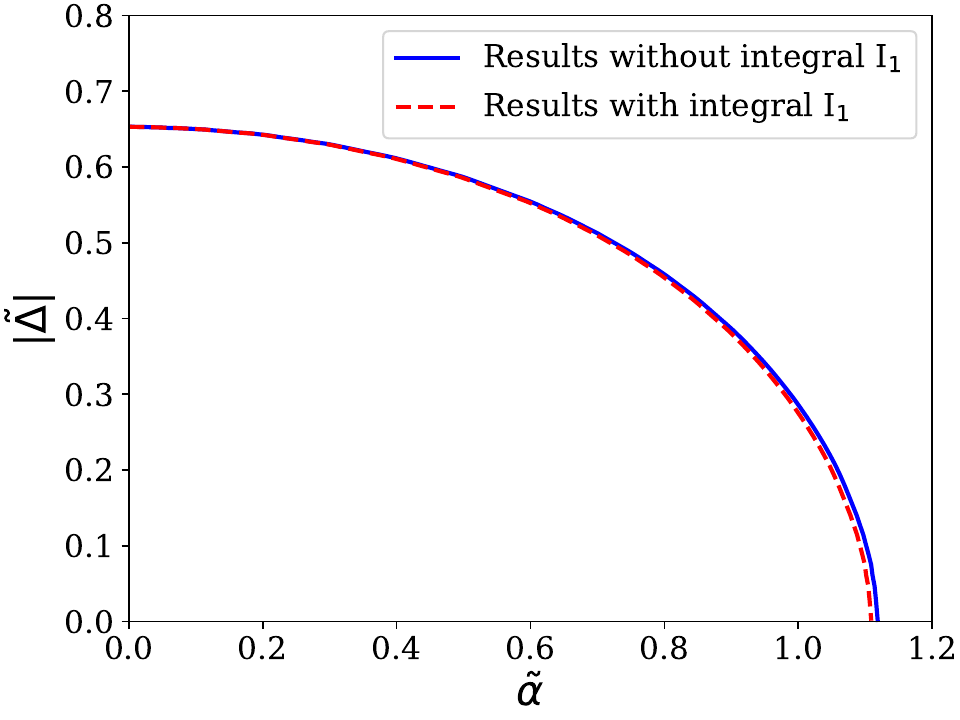}
\caption{The contribution of the $ I_1 $ integral, determined using Eq.~(\ref{eq:i_1}), to the gap equation. The calculations are performed with the values $ \tilde{\mu} = 0.5 $, $ \tilde{\omega}_D = 0.05 $, $ \tilde{J} = 0.4 $, and $ \tilde{\beta} = 0.8 $.}
\label{fig:i_1}
\end{figure}

Since finite temperatures suppress $ \Delta $, the largest contribution of $ I_1 $ to $ \Delta $ occurs at $ T = 0 $. Therefore, we continue our analysis for $ T = 0 $. In this case, using Eqs.~(\ref{eq:gap_main_2}) and (\ref{eq:app2_3}), we can write Eq.~(\ref{eq:app2_1}) as follows
\begin{widetext}
\begin{eqnarray}
    1 &=& \frac{|\lambda| \rho_0^{2D}}{2} \int_{-\pi}^\pi \frac{dk_z}{2\pi} \int_{-\pi}^{\pi} \frac{d\theta}{2\pi} \int_{-\omega_D}^{\omega_D} d\epsilon \left[
    \frac{3}{\sqrt{\epsilon^2+2\eta_1\epsilon+\eta_2}}\right. + \frac{2}{\epsilon \cdot \eta_1 + \eta_2} \left(2 \left(\epsilon - \sqrt{\epsilon^2+2\eta_1\epsilon+\eta_2}\right)\right.\nonumber\\
    &&\left.+\frac{2|\Delta|^2(\sqrt{\epsilon^2+2\eta_1\epsilon+\eta_2} - 1)}{\epsilon \cdot \eta_1 + \eta_2} - \frac{|\Delta|^2}{\sqrt{\epsilon^2+2\eta_1\epsilon+\eta_2}}\right)\Bigg],\nonumber\\
\label{eq:app2_4}    
\end{eqnarray}
\end{widetext}
where $ \epsilon $ and $ \eta_{1,2} $ are defined by Eqs.~(\ref{eq:epsilon}) -- (\ref{eq:lambda2}). By performing the integration with respect to $ \epsilon $ in Eq.~(\ref{eq:app2_4}), we find that
\begin{widetext}
\begin{eqnarray}
    1 
    &=& - \frac{|\lambda| \rho_0^{2D}}{4} \int_{-\pi}^\pi \frac{dk_z}{2\pi} \int_{-\pi}^{\pi} \frac{d\theta}{2\pi} 
    \left[\ln \left(\frac{\eta_3+\eta_4 \sin 2\theta}{4\omega_D^2}\right)\right. + \ln \left(\frac{\eta_3-\eta_4 \sin 2\theta}{4\omega_D^2}\right) 
    - 4 \omega_D^2 \Bigg(\frac{1}{\eta_3+\eta_4 \sin 2\theta }\times \left( \frac{|\Delta|^2}{\eta_3+\eta_4 \sin 2\theta} - 1\right)\nonumber\\
    && + \frac{1}{\eta_3-\eta_4 \sin 2\theta } \times \left( \frac{|\Delta|^2}{\eta_3-\eta_4 \sin 2\theta} - 1\right)\Bigg)\Bigg].
\label{eq:app2_5}  
\end{eqnarray}
\end{widetext}
From Eq.~(\ref{eq:app2_5}), we see that even when the integral $ I_1 $ is present, the contribution of $ J $ to $ \Delta $ at $ T = 0 $ only occurs in the presence of SOI. Finally, using the integrals from Appendix~\ref{app_4} and $\int_{-\pi}^\pi dx \, (a + b \sin x)^{-2} = 2 \pi a (a^2 - b^2)^{-3/2}$, we find that
\begin{widetext}
\begin{eqnarray}
    1 
    &=&- |\lambda| \rho_0^{2D} \int_{-\pi}^\pi \frac{dk_z}{2\pi} \left[
    \ln \left(\frac{\sqrt{|\Delta|^2+2m(\mu-J\cos k_z)(\alpha+\beta)^2}+\sqrt{|\Delta|^2+2m(\mu-J\cos k_z)(\alpha-\beta)^2}}{4\omega_D}\right)\right.\nonumber\\
    && - 2 \omega_D^2\frac{|\Delta|^2 \left(|\Delta|^2+2m(\mu-J\cos k_z)(\alpha^2+\beta^2) \right) - \left(|\Delta|^2+2m(\mu-J\cos k_z)(\alpha^2-\beta^2) \right)^2 + 16 m^2 \mu^2 \alpha^2\beta^2}{\left(|\Delta|^2+2m(\mu-J\cos k_z)(\alpha+\beta)^2\right)^\frac{3}{2}\left(|\Delta|^2+2m(\mu-J\cos k_z)(\alpha-\beta)^2\right)^\frac{3}{2}}\Bigg].
\label{eq:app2_main}
\end{eqnarray}
\end{widetext}
From Eq.~(\ref{eq:app2_main}), it is evident that the contribution of the $ I_1 $ integral to the gap equation calculated using Eq.~(\ref{eq:app2_1}) is a term multiplied by $ \omega_D^2 $. In all practical cases, since the cutoff energy $ \omega_D $ is much smaller than $ \mu $ ($ \omega_D \ll \mu $), this contribution is considered to be very small. This can also be seen from Fig.~\ref{fig:i_1}. In this figure, the solid line represents the gap equation calculated with $ 1 = I_0 $ using Eq.~(\ref{eq:gap_T=0_main}), while the dashed line represents the gap equation calculated with $ 1 = I_0 + I_1 $ using Eq.~(\ref{eq:app2_main}). As shown, the contribution of the $ I_1 $ integral to $ |\Delta| $ is very small. The presence of $ I_1 $ only results in a very slight reduction in the critical value due to SOIs.

\section{\label{app_3} Fermi surface}

\begin{figure*}
\includegraphics[width=0.95\textwidth]{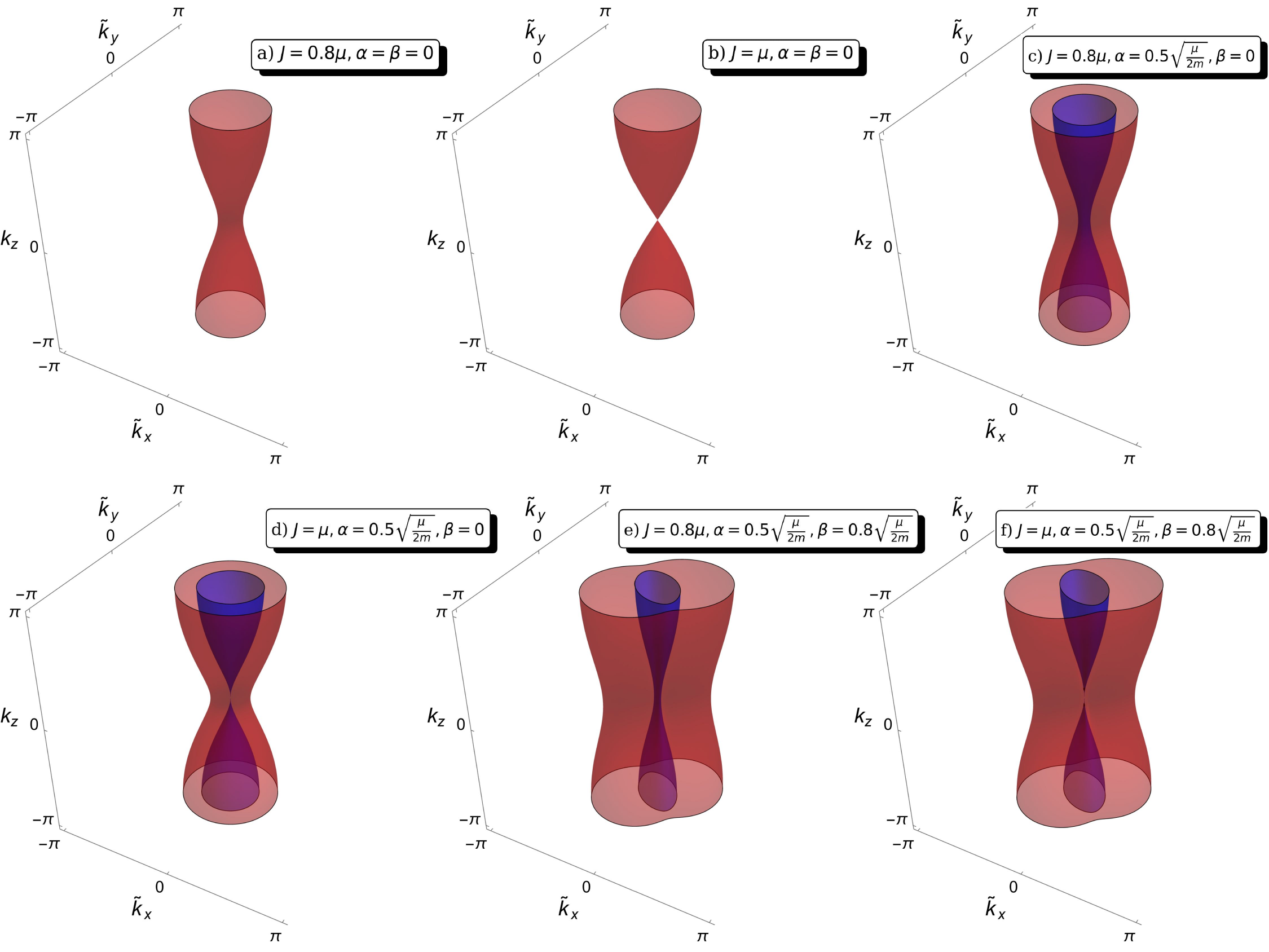}
\caption{Fermi surface of layered superconductors in the presence of Rashba and Dresselhaus SOIs. The dimensionless momenta $\tilde{k}_x$ and $\tilde{k}_y$ are calculated using the formulas $\tilde{k}_{x} = k_{x}/\sqrt{2m\mu}$ and $\tilde{k}_{y} = k_{y}/\sqrt{2m\mu}$.}
\label{fig:Fermi_sur}
\end{figure*}

Based on Eqs.~(\ref{eq:xi}) and (\ref{eq:soc}), the Fermi surface for our model has been computed for different values of $J$, $\alpha$, and $\beta$, as shown in Fig.~\ref{fig:Fermi_sur}. Figs.~\ref{fig:Fermi_sur}a and \ref{fig:Fermi_sur}b consider the cases without SOIs.
For $J=0$, this surface has a uniform cylindrical shape. However, when $J<\mu$, it starts to shrink around $k_z=0$, as seen in Fig.~5a. In the case of $J=\mu$, there is no Fermi surface at $k_z=0$, as illustrated in Fig.~\ref{fig:Fermi_sur}b. Note that for $J>\mu$, $\mathbf{k}=0$ occurs at multiple points just beyond $k_z=0$. 

In the presence of only one SOI, the Fermi surface splits symmetrically into two surfaces. For Rashba SOI, this is shown in Fig.~\ref{fig:Fermi_sur}c for the  $J < \mu$  case and in Fig.~\ref{fig:Fermi_sur}d for the $J = \mu$  case. As seen in Fig.~\ref{fig:Fermi_sur}d, in the presence of SOI, the Fermi surface corresponding to the positive values of $\xi_{SO}$ (depicted in blue in Fig.~\ref{fig:Fermi_sur}d) does not form at $k_z = 0$. Additionally, for the same surface, we observe that the Fermi surface forms for very small $\mathbf{k}$ values in the close vicinity of $k_z = 0$.
The results for  $\alpha \neq 0$ and $\beta \neq 0$ are shown in Figs.~\ref{fig:Fermi_sur}e and \ref{fig:Fermi_sur}f. These results indicate that the presence of both SOIs in the system makes the Fermi surface strongly anisotropic. As seen in Fig.~\ref{fig:Fermi_sur}f, similar to the cases in Fig.~\ref{fig:Fermi_sur}b and Fig.~\ref{fig:Fermi_sur}d, for the Fermi surface corresponding to the positive values of $\xi_{SO}$, $\mathbf{k} = 0$ occurs at $k_z = 0$. For this surface as well, similar to the blue surface in Fig.~\ref{fig:Fermi_sur}d, the presence of SOIs causes the Fermi surface to form at very small $\mathbf{k}$ values near $k_z=0$.

\section{\label{app_4}Integration with respect to the polar angle}

\begin{figure}[H]
\begin{tikzpicture}
    \draw[thick] (3,0) arc[start angle=0, end angle=180, radius=3cm];
    
    \draw[->] (-3.5,0) -- (3.5,0) node[right] {$Re (z)$};
    \draw[->] (0,-0.5) -- (0,3.5) node[above] {$Im (z)$};
    
    \draw[->, line width=0.67pt] (0,0) -- (2.12,2.12) node[midway, above, sloped] {$R$};
    
    \node[below] at (-3,0) {$-R$};
    \node[below] at (3,0) {$R$};
    
    \draw[->, line width=0.67pt] (2.12,2.12) arc[start angle=45, end angle=135, radius=3cm] node[midway, above] {};

    \node at (1.35,2.9) {$\Gamma$};
    \node at (-2.9,3) {$C$};
    \node at (-0.26,1.52) {$i$};

    \fill (0,1.5) circle (2pt);
\end{tikzpicture}
\caption{Semicircular contour for integral (\ref{eq:app_3_2}).}
\label{fig:contour}
\end{figure}
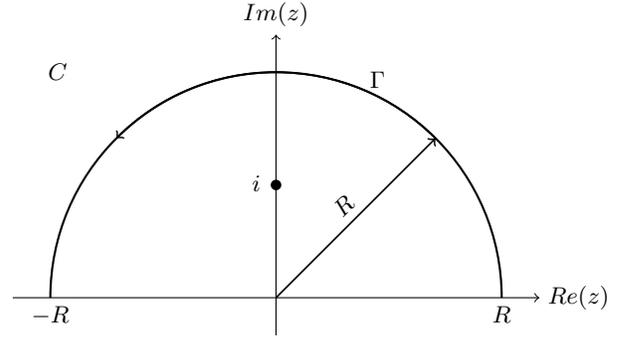

In Eq.~(\ref{eq:gap_T=0}), the integral with respect to the polar angle is given by $\int_{-\pi}^{\pi} d\theta
\ln \left(\eta_3+\eta_4 \sin 2\theta\right)$, where $\eta_3$ and $\eta_4$ are expressed in Eqs.~(\ref{eq:lambda3}) and (\ref{eq:lambda4}), respectively. To evaluate this integral, we first make the substitution $\varphi = 2\theta - \frac{3\pi}{2}$, and then substitute $z = \tan\varphi$,
\begin{eqnarray}
    \int_{-\pi}^{\pi} d\theta
    \ln \left(\eta_3+\eta_4 \sin 2\theta\right) = \int_{-\infty}^{\infty}\frac{dz}{1+z^2} \nonumber\\
    \times \ln \left(\frac{\eta_3^2-\eta_4^2+\eta_3^2z^2}{1+z^2}\right).
\label{eq:app_3_1}    
\end{eqnarray}
To evaluate the integral in Eq.~(\ref{eq:app_3_1}), we consider the complex integral 
$\oint_C \frac{dz}{1+z^2} \ln\left(i\sqrt{\eta_3^2-\eta_4^2}+\eta_3z\right)$. 
As shown in Fig.~\ref{fig:contour}, we choose an analytical function in the upper half-plane. Then
\begin{eqnarray}
    &\oint_C \frac{dz}{1+z^2} \ln\left(i\sqrt{\eta_3^2-\eta_4^2}+\eta_3z\right) 
    = 2\pi i\nonumber\\ 
    &\times \lim_{z\rightarrow i} \frac{\ln\left(i\sqrt{\eta_3^2-\eta_4^2}+\eta_3z\right)}{z+i}\nonumber\\
    & = \pi \ln\left(\eta_3+\sqrt{\eta_3^2-\eta_4^2}\right) + \frac{i}{2}\pi^2.
\label{eq:app_3_2}  
\end{eqnarray}
On the other hand, for the contour shown in Fig.~\ref{fig:contour}, we can write
\begin{eqnarray}
    &\oint_C \frac{dz}{1+z^2} \ln\left(i\sqrt{\eta_3^2-\eta_4^2}+\eta_3z\right) \nonumber\\
    &= \int_{-R}^R \frac{dz}{1+z^2} \ln\left(i\sqrt{\eta_3^2-\eta_4^2}+\eta_3z\right) \nonumber\\
    &+ \int_\Gamma \frac{dz}{1+z^2} \ln\left(i\sqrt{\eta_3^2-\eta_4^2}+\eta_3z\right),
\label{eq:app_3_3}  
\end{eqnarray}
where the integral along $\Gamma$ on the right-hand side is zero. At the same time,
\begin{eqnarray}
    &\int_{-R}^R \frac{dz}{1+z^2} \ln\left(i\sqrt{\eta_3^2-\eta_4^2}+\eta_3z\right) \nonumber\\
    &=\int_{0}^R \frac{dz}{1+z^2} \ln\left(\eta_3^2-\eta_4^2+\eta_3^2z\right) \nonumber\\
    &+ \int_{0}^R \frac{dz}{1+z^2} i\pi
\label{eq:app_3_4} 
\end{eqnarray}
holds. Using Eqs.~(\ref{eq:app_3_3}) and (\ref{eq:app_3_4}), we find from Eq.~(\ref{eq:app_3_2}) that
\begin{equation}
    \int_{-\infty}^\infty \frac{dz}{1+z^2} \ln\left(\eta_3^2-\eta_4^2+\eta_3^2z\right) = 2\pi \ln\left(\eta_3+\sqrt{\eta_3^2-\eta_4^2}\right).
\end{equation}
Thus, for the integration with respect to the polar angle in the gap equation at $T = 0$, we find that
\begin{equation}
    \int_{-\pi}^{\pi} d\theta
    \ln \left(\eta_3+\eta_4 \sin 2\theta\right) = 2\pi \ln\left(\frac{\eta_3+\sqrt{\eta_3^2-\eta_4^2}}{2}\right).
\label{eq:app_3_main}    
\end{equation}

\newpage
\nocite{*}
\bibliography{biblio}

\end{document}